\documentclass[3p,times]{elsarticle}

\usepackage{lineno,hyperref}
\usepackage{floatrow}

\modulolinenumbers[5]

\journal{Annals of Physics}

\biboptions{sort&compress}

\usepackage{amssymb,amsmath,textcomp}
\usepackage{float}
\usepackage{bm}

\usepackage{graphicx}
\usepackage{epsfig,epsf}
\usepackage{epstopdf}

\usepackage{makeidx}

\usepackage{color}
\usepackage{hyperref}
\usepackage{caption}
\usepackage{subcaption}

\usepackage{setspace}

\makeindex
\newcommand{\beq}{\begin{eqnarray}}
	\newcommand{\eeq}{\end{eqnarray}}

\def\be{\begin{equation}}
	\def\ee{\end{equation}}
\def\la{\langle}
\def\ra{\rangle}

\newcommand\eq[1]{Eq.~(\ref{#1})}

\numberwithin{equation}{section}

\begin{document}

\begin{frontmatter}

	\title{\centering{Casimir versus Helmholtz forces in the Gaussian model: exact results for Dirichlet--Dirichlet,  Neumann--Dirichlet, Neumann--Neumann, and periodic boundary conditions}}
	\author[DMD,DMD2,DMD3,JR]{D. M. Dantchev\corref{CorrespondingAuthor}}
	\ead{daniel@imbm.bas.bg}
	%\ead[url]{www.apmaths.uwo.ca/people/vmiransky.shtml}
	%\ead[url]{http://shovkovy.faculty.asu.edu}
	\author[JR]{J. Rudnick}
	\ead{jrudnickucla@gmail.com}
	
	\address[DMD]{Institute of
		Mechanics, Bulgarian Academy of Sciences, Academic Georgy Bonchev St. building 4,
		1113 Sofia, Bulgaria}
		\address[DMD2]{Center of Competence for Mechatronics and Clean Technologies “Mechatronics, Innovation, Robotics, Automation and Clean Technologies” - MIRACle, “Acad. G.		Bontchev” Str. 4, 1113 Sofia, Bulgaria}
		\address[DMD3]{Max-Planck-Institut f\"{u}r Intelligente Systeme, Heisenbergstrasse 3, D-70569 Stuttgart, Germany}
	\address[JR]{Department of Physics and Astronomy, University of California, Los Angeles, CA 90095}
	
	\cortext[CorrespondingAuthor]{Corresponding author}

	
	\begin{abstract}
		We present results and compare the behavior of two fluctuation-induced forces pertinent for their corresponding ensembles: the critical Casimir force in the grand canonical (fixed external field $h$) one and the critical Helmholtz force in the canonical (fixed average value of the order parameter $m$) one. We do so by deriving exact results for their behavior near the bulk critical point at $T=T_c$ in the three-dimensional Gaussian model. We consider  Dirichlet-Dirichlet,  Neumann-Dirichlet, Neumann-Neumann,  and periodic boundary conditions. For every boundary condition examined, we confirm that both forces follow a finite-size scaling. We find that  for Dirichlet-Dirichlet and Neumann-Dirichlet boundary conditions the Casimir and the Helmholtz force differ from each other. For Dirichlet-Dirichlet boundary conditions the Casimir force is always attractive, while the Helmholtz force can be both attractive and repulsive as a function of $T$ and $m$. For Neumann-Dirichlet boundary conditions the Casimir force changes sign from repulsive to attractive with increase of $h$, while the Helmholtz force stays always repulsive.  Under periodic and Neumann-Neumann boundary conditions the Casimir force and the Helmholtz force coincide - the first does not depend on $h$, while the latter does not depend on $m$; they are always attractive.
	\end{abstract}
	
	\date{\today}
	
\end{frontmatter}

\section{Introduction}

Thermodynamic ensembles are generally equivalent for infinite (bulk) systems, but this equivalence does not hold for finite systems. In this article, we will discuss two fluctuation-induced forces: the Casimir force, which is relevant to the grand canonical ensemble (GCE), and the Helmholtz force, associated with the canonical ensemble (CE). We will analyze these forces as a function of temperature \( T \) and external magnetic field \( h \) for the Casimir force, and as a function of temperature \( T \) and average magnetization \( m \) for the Helmholtz force. The explicit calculations will be carried out using the Gaussian model, a fundamental model in statistical mechanics that often serves as the starting point for more advanced methods, such as normalization group calculations.  

 Let us envisage a system with a film geometry $\infty^{d-1}\times L$, and with boundary conditions $\tau$ imposed along the spatial direction of finite extent $L$.  Take ${\cal F}_{ {\rm total}}^{(\tau)}$ to be the total free energy of such a system within the GCE. Then, if   $f^{(\tau)}(T,h,L)\equiv \lim_{A\to\infty}{\cal F}_{ {\rm total}}^{(\tau)}/A$  is the free energy per area $A$ of the system, one can define the Casimir force for critical systems in the grand-canonical $(T-h)$-ensemble, see, e.g. Ref. \cite{Dantchev2023}, as: 
 \begin{equation}
 	\label{CasDef}
 	\beta F_{\rm Casimir}^{(\tau)}(L,T,h)\equiv- \frac{\partial}{\partial L}f_{\rm ex}^{(\tau)}(L,T,h)
 \end{equation}
 where
 \begin{equation}
 	\label{excess_free_energy_definition}
 	f_{\rm ex}^{(\tau)}(L,T,h) \equiv f^{(\tau)}(L,T,h)-L f_{\rm bulk}(T,h)
 \end{equation}
 is the so-called excess, over the bulk $f_{\rm bulk}(T,h)$ one, free energy per area and per $\beta^{-1}=k_B T$. 
 
 Along these lines we define the corresponding Helmholtz  fluctuation induced force in canonical $(T-m)$-ensemble:
 \begin{equation}
 	\label{HelmDef}
 	\beta F_{\rm Helmholtz}^{(\tau)}(L,T,m)\equiv- \frac{\partial}{\partial L}a_{\rm ex}^{(\tau)}(L,T,m)
 \end{equation}
 where
 \begin{equation}
 	\label{excess_free_energy_definition_M}
 	a_{\rm ex}^{(\tau)}(L,T,m) \equiv a^{(\tau)}(L,T,m)-L a_{\rm bulk}(T,m),
 \end{equation}
 with $m$ being the average magnetization in the system, and $a_{\rm bulk}$ is the Helmholtz free energy density of the ``bulk'' system. 
 
 In the remainder we will take $L=N a$, where $N$ is an integer number, and for simplicity we 
 set $a=1$, i.e., all lengths will be measured in units of the lattice spacing $a$. Note that neither $f_{\rm bulk}$, nor $a_{\rm bulk}$ depend on the boundary conditions. 
 We will show that the Helmholtz fluctuation induced force  has a behavior very different from that of the Casimir force.
 
 According to  finite-size scaling theory \cite{Bb83,P90,ET87,BDT2000,DD2022} near the critical point of of the bulk system the singular \ the excess free energies $f_{\rm ex}^{(\tau)}$ and $a_{\rm ex}^{(\tau)}$ (in units of $k_B T$) take the form
 \begin{equation}
 	\label{eq:gibbs-free-energy}
 	f_{\rm ex}^{(\tau)}(L,T,h) =L^{-(d-1)} X_{\rm ex, GC}^{(\tau)} (x_t,x_h) + \mbox{corrections}
 \end{equation}
 and 
 \begin{equation}
 	\label{eq:helmholtz-free-energy}
 	a_{\rm ex}^{(\tau)}(L,T,h) =L^{-(d-1)} X_{\rm ex, C}^{(\tau)} (x_t,x_m) + \mbox{corrections},
 \end{equation}
 where 
 \begin{equation}
 	\label{eq:scaling-variables}
 	x_t=a_t t L^{1/\nu}, \quad x_h=a_h h L^{\Delta/\nu}  \quad \mbox{and} \quad x_m=a_m m L^{\beta/\nu}, 
\end{equation}
where $a_t$, $a_h$ and $a_m$ are non-universal constants, $X_{\rm ex, GC}$ and $X_{\rm ex, C}$ are  universal scaling functions, and $t=(T-T_c)/T_c$, where $T$ has the meaning of the temperature of the system, and $T_c$ is its bulk transition temperature.  For the susceptibility one has, correspondingly 
 \begin{equation}
 	\label{eq:susceptibility-scaling}
 	\chi_L^{(\zeta)}(t,h)=a_h^2 L^{\gamma/\nu} X_\chi^{(\tau)}(a_t t L^{1/\nu}, a_h h L^{\Delta/\nu}). 
 \end{equation}

Let us emphasize, that  in any ensemble --- not only in  the grand canonical and canonical, but also, say in  the micro-canonical --- one can define a thermodynamic fluctuation induced force which is specific for that ensemble. By their definition, these ensembles correspond to quite different physical conditions. Thus,  it is reasonable to
expect that the behavior of these forces differ from one ensemble to another. The precise
behavior of these forces, however, has not yet been the object of thorough and systematic
study. 

Without a doubt, the most well-known fluctuation induced force is the Casimir force which is due to the quantum, or temperature-driven, fluctuation of the electromagnetic field between polarizable objects. This quantum-mechanical Casimir effect is the subject of multiple review articles --- see, e.g., \cite{BMM2001,M2001,BKMM2009,RCJ2011,WKD2021} and literature cited therein. It is named after the Dutch physicist H. B. Casimir who first realized that in the case  of two perfectly-conducting, uncharged, and smooth plates parallel to each other in vacuum, at $T=0$  these fluctuations lead to an \textit{attractive} force between them \cite{C48}. Thirty years later, Fisher and De Gennes \cite{FG78} showed that a very similar effect exists in critical fluids, today known as critical Casimir effect. It is a force  that arises in systems, theoretically described by the grand canonical ensemble. A summary of the results available for this effect can be found in recent reviews \cite{MD2018,DD2022,Gambassi2024,Dantchev2024a}. The description of the critical Casimir effect is based on  finite-size scaling theory \cite{Ba83,P90,BDT2000}. We note that the critical Casimir effect has been observed experimentally \cite{GC99,GC2002,GSGC2006,HHGDB2008,Schmidt2022}. The notion of the Helmholtz fluctuation induced force has been introduced relatively recently, in Ref. \cite{DR2022}. Such a force is pertinent to systems with fixed average value of the magnetization $m$, i.e., for the canonical ensemble. It has been investigated for the Ising chain with periodic, antiperiodic and Dirichlet boundary conditions \cite{DR2022,Dantchev2023b,Dantchev2024b,Dantchev2025a} and in the Nagel -- Kardar model \cite{Dantchev2024c,Dantchev2025b}. The behavior of the force turns out to be quite different from that  of the Casimir force. More specifically, it changes it sign as a function of the temperature, being both attractive and repulsive, while the Casimir force is always attractive under the very same boundary conditions. Currently, the only existing review on Helmholtz forces is given in Ref. \cite{Dantchev2024a}.

In the current article we complement the existing results for the Ising and Nagel-Kardar model with another basic model of statistical mechanics --- the Gaussian model.  The issue of the ensemble dependence of the fluctuation induced forces is, however, by no means limited to specific models  and can be addressed, in principle, in \textit{any} model of interest. The analysis reported here can also be viewed as a useful addition to approaches to fluctuation-induced forces in the fixed-$m$ ensemble based on the Ginsburg-Landau-Wilson Hamiltonian \cite{RSVG2019,GVGD2016,GGD2017} in which the authors studied, in a mean-field approximation, the ``canonical Casimir force'' which, in fact, coincides with the Helmholtz force defined above.

From \eq{eq:gibbs-free-energy} and \eq{eq:helmholtz-free-energy} for the scaling functions of the Casimir and Helmholtz forces we derive
\begin{equation}
	\label{eq:connection-Casimir}
	X_{\rm Casimir}^{(\tau)}(x_t,x_h)=(d-1) X_{\rm ex,\; GC}^{(\tau)}(x_t,x_h)-\frac{1}{\nu} x_t \frac{\partial}{\partial x_t}X_{\rm ex\; GC}^{(\tau)}(x_t,x_h)-\frac{\Delta}{\nu}\frac{\partial}{\partial x_h}X_{\rm ex\; GC}^{(\tau)}(x_t,x_h)
\end{equation}
and
\begin{equation}
	\label{eq:connection-Helmholtz}
	X_{\rm Helmholtz}^{(\tau)}(x_t,x_m)=(d-1) X_{\rm ex,\; C}^{(\tau)}(x_t,x_m)-\frac{1}{\nu} x_t \frac{\partial}{\partial x_t}X_{\rm ex\; C}^{(\tau)}(x_t,x_m)-\frac{\beta}{\nu}\frac{\partial}{\partial x_m}X_{\rm ex\; C}^{(\tau)}(x_t,x_m)
\end{equation}

The critical behavior of the bulk Gaussian model is well known. Its critical exponents for $2<d\le 4$ are \cite{G92,BDT2000}
	\begin{eqnarray}
		\label{eq:critical-exponents}
		\alpha &=& 2-d/2, \quad  \nu=1/2, \quad \eta=0, \quad \beta =(d-2)/4, \nonumber \\
		\delta &=& \frac{d+2}{d-2}, \quad \gamma=1, \quad \Delta=(1-d/2)/2.
\end{eqnarray}

The structure of the article is as follows. The definitions of the Gaussian model on a lattice and in continuum are presented in Secs. \ref{sec:Gaussian-lattice} and \ref{sec:continuum}. This is done for both grand canonical and canonical ensembles. We consider Dirichlet -- Dirichlet, Dirichlet -- Neumann, Neumann -- Neumann, and periodic boundary conditions. The the results for: \textit{i)} the Dirichlet-Dirichlet boundary conditions  are reported in Secs. \ref{sec:Dirichlet-Dirichlet-bc} and \ref{sec:Dirichlet-Dirichlet}; for \textit{ii)} the Neumann-Dirichlet ones  --- in Sec. \ref{sec:Neumann-Dirichlet-bc}, \ref{sec:ND} and \ref{app:Neumann-Dirichlet}; for \textit{iii)} Neumann-Neumann ones - in Secs. \ref{sec:Neumann-Neumann-bc}, \ref{sec:NN} and \ref{app:Neumann-Neumann}; \textit{ iv)} for periodic ones in Sec: \ref{sec:per-boundary-conditions} and \ref{sec:periodic}. The most of the technical details are given in the appendixes mentioned. The article ends with Sec. "Concluding remarks and discussion".

\section{On the lattice Gaussian model in grand canonical and canonical al ensembles}
\label{sec:Gaussian-lattice}

We consider a ferromagnetic model with nearest-neighbor interactions on a fully finite $d$-dimensional hypercubic lattice $\Lambda \in Z^d$ of $|\Lambda|$ sites. 	Let us take $\Lambda \in Z^d$ to be the parallelepiped
$\Lambda ={\mathcal L}_1 \times \dots \times {\mathcal L}_d$, where $\times$
denotes the direct (Cartesian) product of the finite sets ${\mathcal L}_{\nu} =
\{1,\dots ,L_{\nu}\}$. 

It is
convenient to consider the configuration space $\Omega_{\Lambda}=R^{|\Lambda|}$
as an Euclidean vector space in which each configuration is
represented by a column-vector $S_{\Lambda}$ with components labeled according
to the lexicographic order of the set $\{ \mathbf r=(r_{1}, \cdots,r_{d}) \in \Lambda \}$.
Let $S_{\Lambda}^{\dagger}$ be the corresponding transposed row-vector and let
the dot ($\cdot$) denote matrix multiplication. Then, for given boundary
conditions $\tau = (\tau_{1}, \cdots,\tau_{d})$, specified for each pair of
opposite faces of $\Lambda$ by some $\tau_{\nu}$ takes the form
\begin{equation}
	\beta {\mathcal H}^{(\tau)}_{\Lambda}(S_{\Lambda}|K) = -
	{1\over 2}\; K\; S_{\Lambda}^{\dagger}\cdot I_{\Lambda}^{(\tau)}\cdot S_{\Lambda}.
	\label{matrintH}
\end{equation}
Here $K=\beta J$, where $J$ is the interaction constant (to be set to $J=1$ in the remainder), and the $|\Lambda|\times |\Lambda|$ interaction matrix $I^{(\tau)}_{\Lambda}$
can be written as
\begin{equation}
	I^{(\tau)}_{\Lambda} = (\Delta_{1}^{(\tau_{1})}+2\: E_{1}) \times
	\cdots \times
	(\Delta_{d}^{(\tau_{d})}+2\: E_{d}),
	\label{Q-as-L}
\end{equation}
where $\Delta_{\nu}^{(\tau_{\nu})}$ is the one-dimensional discrete Laplacian
defined on the finite chain ${\mathcal L}_{\nu}$ under boundary conditions
$\tau_{\nu}$, and $E_{\nu}$ is the $L_{\nu}\times L_{\nu}$ unit matrix. 

By using the results of  \cite[Chapter 7]{BDT2000}, see also Ref. \cite{Dantchev2025}, we can write down the
eigenfunctions of the interaction matrix (\ref{Q-as-L}) in the form
\begin{equation}
	u_{\Lambda}^{(\tau)}({\mathbf r},{\mathbf k}) = u_{L_{1}}^{(\tau_{1})}
	(r_{1},k_{1}) \:\cdots \:
	u_{L_{d}}^{(\tau_{d})}(r_{d},k_{d}),\hspace{1cm} \mathbf k =(k_1,\cdots,k_d)\in \Lambda,
	\label{general-eigenvalues}
\end{equation}
and obtain the corresponding eigenvalues of it
\begin{equation}
	\mu_{\Lambda}^{(\tau)}({\mathbf k}) =
	2\sum_{\nu = 1}^{d}\cos \varphi_{_{L_{\nu}}}^{(\tau_{\nu})}(k_{\nu}),
	\hspace{1cm} {\mathbf k}\in \Lambda .
	\label{general-eigenvalues}
\end{equation}
Obviously, $\max_{{\mathbf k} \in \Lambda} \mu_{\Lambda}^{(\tau)}({\mathbf k}) =2d$. Note that the interaction Hamiltonian (\ref{matrintH}) has negative eigenvalues,
which necessitates the inclusion of a positive-definite quadratic form in
the Gibbs exponent to ensure the existence of the corresponding partition function. Thus, we consider the Hamiltonian 
\begin{equation}
	\beta {\mathcal H}^{(\tau)}_{\Lambda}(S_{\Lambda}|\beta;s) = -
	{1\over 2} \; K \; S_{\Lambda}^{\dagger} \cdot I_{\Lambda}^{(\tau)} \cdot
	S_{\Lambda} + s\; S_{\Lambda}^{\dag} \cdot S_{\Lambda}.
	\label{eq:the-Hamiltonian}
\end{equation}

In order to ensure the existence of the partition function, all the eigenvalues
$-{1\over 2} K \mu_{\Lambda}^{(\tau)}({\mathbf k})+s$, ${\mathbf k}\in \Lambda$,
of the quadratic form in $\beta {\mathcal H}^{(\tau)}_{\Lambda}
(S_{\Lambda}|\beta;s)$, 
ought to be positive. Hence, the field $s^{(\tau)}$ must satisfy the inequality
\begin{equation}
	s >{1\over 2} K \max_{{\mathbf k} \in \Lambda} \mu_{\Lambda}^{(\tau)}
	({\mathbf k}) \equiv {1\over 2} K \mu_{\Lambda}^{(\tau)}({\mathbf k}_{0}),
	\label{DBA12.21}
\end{equation}
with
\begin{equation}
	\label{eq:critical-temperature}
	K_{c,L}={1\over 2}\mu_{\Lambda}^{(\tau)}({\mathbf k}_{0})
\end{equation}
defining the critical temperature of the \textit{finite} system. Since, as stated above $\max_{{\mathbf k} \in \Lambda} \mu_{\Lambda}^{(\tau)}({\mathbf k}) =2d$, it is clear that for the infinite system 
\begin{equation}
	\label{eq:bulk-critical-temperature}
	K_{c}=d. 
\end{equation}

\subsection{The behavior of the Gibbs free energy in lattice Gaussian model}
According to Refs. \cite[Chapter 7]{BDT2000}, and Ref. \cite{Dantchev2025}, one has that the Gibbs free energy density of a \textit{finite} system in a region $\Lambda$, obtained from $Z_{GM}^{GC}$, is
\begin{eqnarray}
	\beta f_{\Lambda}^{(\tau)}(K,h_{\Lambda}) = {1\over 2}\left\{\ln(K/2\pi)
	+\, {\mathcal U}_{\Lambda}^{(\tau)}(K,s) -
	P_{\Lambda}^{(\tau)}(K,h_{\Lambda},s) \right\}.
	\label{eq:free-energy-density-finite-region}
\end{eqnarray}

In \eq{eq:free-energy-density-finite-region} the first term does not depend on the size of the system, i.e., they are the same in both finite and infinite systems. However, the other two terms do depend on the size of the system. The function ${\mathcal U}_{\Lambda}^{(\tau)}(K,s)$ is due to the spin-spin interaction (and will be called ``interaction term''); it depends on $s$, but does not depend on $h$. It is equal to
\begin{equation}
	{\mathcal U}_{\Lambda}^{(\tau)}(K,s) = |\Lambda|^{-1}\sum_{{\mathbf k}\in
		\Lambda} \ln \left[\frac{2s}{K} - \mu_{\Lambda}^{(\tau)}({\mathbf k})\right],
	\label{eq:interaction-term-finite-region}
\end{equation}
which is obtained after performing the corresponding Gaussian integrals in the free energy of the finite system. Here   $h_{\Lambda}= \{h({\mathbf r}),\,{\mathbf r}\in \Lambda\}$
is a column-vector representing (in units of $k_B T$) the inhomogeneous
magnetic field configuration acting upon the system, and $h_{\Lambda}^{\dagger}$ is the transposed row-vector.  The dependence of the free energy on the field variables $h$ is given by the "field term"  
\begin{equation}
	P_{\Lambda}^{(\tau)}(K,h_{\Lambda};s) = {1\over K |\Lambda|}
	\sum_{{\mathbf k}\in \Lambda} {[\hat{h}_{\Lambda}^{(\tau)}
		({\mathbf k})]^{2} \over 2s/K - \mu_{\Lambda}^{(\tau)}({\mathbf k})}.
	\label{eq:the-field-term}
\end{equation}
Here $\hat{h}_{\Lambda}^{(\tau)}
({\mathbf k})$ denotes the projection of the magnetic field configuration
$h_{\Lambda}$ on the eigenfunction $\{\bar{u}_{\Lambda}^{(\tau)}({\mathbf r},
{\mathbf k}),$ ${\mathbf k}\in \Lambda \}$ ( by $\bar{u}$ we denote
the complex conjugate of $u\in {\mathbf C}$):
\begin{equation}
	\hat{h}_{\Lambda}^{(\tau)}({\mathbf k}) = \sum_{{\mathbf r}\in \Lambda}
	h({\mathbf r})\bar{u}_{\Lambda}^{(\tau)}({\mathbf r},{\mathbf k}).
	\label{eq:h-vector}
\end{equation}
When $h({\mathbf r})=i\omega$ the above expression simplifies to 
\begin{equation}
	\hat{h}_{\Lambda}^{(\tau)}({\mathbf k}) = i \omega \sum_{{\mathbf r}\in \Lambda}
	\bar{u}_{\Lambda}^{(\tau)}({\mathbf r},{\mathbf k}).
	\label{eq:h-vector}
\end{equation}

Defining $K_c$ so, that 
\begin{equation}
	\label{eq:critical-temperature}
	\frac{2s}{K}=2d \frac{K_c}{K},
\end{equation}
the above expression can be rewritten in the form
\begin{equation}
	{\mathcal U}_{\Lambda}^{(\tau)}(K) = |\Lambda|^{-1}\sum_{{\mathbf k}\in
		\Lambda} \ln \left[2d (K_c/K-1) +2d- \mu_{\Lambda}^{(\tau)}({\mathbf k})\right].
	\label{eq:interaction-term-finite-region-s}
\end{equation}
The behavior of the interaction term in the bulk system for $d=3$ is 
\begin{equation}
	\label{eq:bulk-interaction-term}
	{\mathcal U}_{\infty,3}(K)=	{\mathcal V}_3\left[6 (K_c/K-1)\right],
\end{equation}
where 
\begin{equation}
	{\mathcal V}_d(z) :={1\over (2\pi)^d} \int_{-\pi}^{\pi} {\mathrm d}\theta_1
	\cdots \int_{-\pi}^{\pi}{\mathrm d} \theta_d \ln \left[z+ 2\sum_{\nu =1}^d
	(1-\cos\theta_{\nu})\right].
	\label{eq:V-function}
\end{equation}
For the field term one has 
\begin{equation}
	\label{eq:bulk-limit}
P_{\infty}(\beta,h)= \frac{h^2}{ 6\beta (K_c/K-1) }, 
\end{equation}

\subsection{The behavior of the Helmholtz free energy in lattice Gaussian model}
In order to calculate the behavior of the system with a fixed total value $M$ in the system we have to calculate the partition function in the canonical ensemble
\begin{equation}
	\label{eq:partition-function}
	Z_{GM}^C(K,M;s)=\int_{{\Omega_\Lambda}} \exp \left[-\beta {\mathcal H}^{(\tau)}_{\Lambda}(S_{\Lambda}|\beta;s) \right] \delta \left[\sum _{i\in \Lambda} S_i-M \right] \prod_{i\in \Lambda} dS_i.
\end{equation}
Using for the delta function the representation 
\begin{equation}
	\label{eq:delta-function}
	\delta(t-u)=\frac{1}{2 \pi} \int_{-\infty}^{\infty} \exp\left[i \omega (t-u)\right] d\omega
\end{equation}
we arrive at
	\begin{eqnarray}
		\label{eq:partition-function-M}
		Z_{GM}^C(K,M;s)&=&\frac{1}{2 \pi} \int_{-\infty}^{\infty} \exp\left[-i \omega M \right] \left\{\int_{{\Omega_\Lambda}} \exp \left[\frac{1}{2}K \sum_{\la i,j\ra \in \Lambda} S_i S_j + i \omega \sum_{i\in \Lambda }S_i - s \sum_{i\in \Lambda }S_i^2 \right]  \prod_{i\in \Lambda} dS_i \right\}d\omega\\
		&=& \frac{1}{2 \pi} \int_{-\infty}^{\infty} \exp\left[-i \omega M \right] \; Z_{GM}^{GC}(\beta,i \omega;s) d \omega ,
	\end{eqnarray}
where $\la \cdots \ra $ means summation over nearest neighbors, and $Z_{GM}^{GC}(\beta,i \omega;s)$ is the partition function of the system in grand canonical ensemble with homogeneous imaginary magnetic field $i \omega$. 
From \eq{eq:partition-function-M} for the Helmholtz free energy $\beta a_{\Lambda}^{(\tau)}(K,M) =-\ln Z_{GM}^C(K,M;s) $ one has 
	\begin{eqnarray}
		\beta a_{\Lambda}^{(\tau)}(K,M) &=& {1\over 2}\left\{\ln(K/2\pi)
		+\, {\mathcal U}_{\Lambda}^{(\tau)}(K,s)\right\} - \frac{1}{|\Lambda|}\ln \frac{1}{2 \pi} \int_{-\infty}^{\infty} \exp\left[-i \omega M \right] 
		\exp\left[\frac{1}{2}|\Lambda| \; P_{\Lambda}^{(\tau)}(K,i \omega;s)\right] d\omega \nonumber \\
		&=& {1\over 2}\left\{\ln(K/2\pi)
		+\, {\mathcal U}_{\Lambda}^{(\tau)}(K,s)\right\}- \frac{1}{|\Lambda|}\ln \frac{1}{2 \pi} \int_{-\infty}^{\infty} \exp\left[-i \omega M -\frac{\omega^2}{2K} 
		\sum_{{\mathbf k}\in \Lambda} {[\sum_{{\mathbf r}\in \Lambda}
			\bar{u}_{\Lambda}^{(\tau)}({\mathbf r},{\mathbf k})]^{2} \over 2s/K - \mu_{\Lambda}^{(\tau)}({\mathbf k})}\right] d\omega \nonumber\\
		&=& {1\over 2}\left\{\ln(K/2\pi)
		+\, {\mathcal U}_{\Lambda}^{(\tau)}(K,s)\right\}- \frac{1}{|\Lambda|}\ln \frac{1}{\sqrt{\frac{2\pi}{K} 
				\sum_{{\mathbf k}\in \Lambda} {[\sum_{{\mathbf r}\in \Lambda}
					\bar{u}_{\Lambda}^{(\tau)}({\mathbf r},{\mathbf k})]^{2} \over 2s/K - \mu_{\Lambda}^{(\tau)}({\mathbf k})}}}\exp \left \{-\frac{M^2}{\frac{2}{K} 
			\sum_{{\mathbf k}\in \Lambda} {[\sum_{{\mathbf r}\in \Lambda}
				\bar{u}_{\Lambda}^{(\tau)}({\mathbf r},{\mathbf k})]^{2} \over 2s/K - \mu_{\Lambda}^{(\tau)}({\mathbf k})}} \right \} \nonumber \\
		&=& {1\over 2}\left\{\ln(K/2\pi)
		+\, {\mathcal U}_{\Lambda}^{(\tau)}(K,s)+ \left[ \frac{1}{|\Lambda|} \ln \frac{2\pi}{K} \sum_{{\mathbf k}\in \Lambda} {[\sum_{{\mathbf r}\in \Lambda}
			\bar{u}_{\Lambda}^{(\tau)}({\mathbf r},{\mathbf k})]^{2} \over 2s/K - \mu_{\Lambda}^{(\tau)}({\mathbf k})}\right]+ \frac{ K M^2/|\Lambda|}{
			\sum_{{\mathbf k}\in \Lambda} {[\sum_{{\mathbf r}\in \Lambda}
				\bar{u}_{\Lambda}^{(\tau)}({\mathbf r},{\mathbf k})]^{2} \over 2s/K - \mu_{\Lambda}^{(\tau)}({\mathbf k})}} \right\} \nonumber \\
		&=& {1\over 2}\left\{\ln(K/2\pi)
		+\, {\mathcal U}_{\Lambda}^{(\tau)}(K,s)+ Q_{\Lambda}^{(\tau)}(K,M;s)\right\}.
		\label{eq:free-energy-density-finite-region-M} 
	\end{eqnarray}
where 
\begin{eqnarray}
	\label{eq:Q1}
	Q_{\Lambda}^{(\tau)}(K,M;s)=\left[ \frac{1}{|\Lambda|} \ln \frac{2\pi}{K} \sum_{{\mathbf k}\in \Lambda} {[\sum_{{\mathbf r}\in \Lambda}
		\bar{u}_{\Lambda}^{(\tau)}({\mathbf r},{\mathbf k})]^{2} \over 2s/K - \mu_{\Lambda}^{(\tau)}({\mathbf k})}\right]+ \frac{ K M^2/|\Lambda|}{
		\sum_{{\mathbf k}\in \Lambda} {[\sum_{{\mathbf r}\in \Lambda}
			\bar{u}_{\Lambda}^{(\tau)}({\mathbf r},{\mathbf k})]^{2} \over 2s/K - \mu_{\Lambda}^{(\tau)}({\mathbf k})}}.
\end{eqnarray}
	
\subsection{Details for the boundary conditions}
\label{sec:laatice-bc}

Using the notations of \cite[Chapter 7]{BDT2000}, 
below we give a list of the complete sets of orthonormal eigenfunctions,
$\{u_{L}^{(\tau)}(r,k)$, $k=1,\dots ,L\}$, of the one-dimensional discrete
Laplacian under the  Neumann - Dirichlet (ND) boundary conditions:
\begin{itemize}
	\item periodic (p) boundary conditions
	\begin{equation}
		u_{L}^{(p)}(r,k) =L^{-1/2} \exp [-{\mathrm i} r\varphi_{_{L}}^{(p)}(k)];
		\label{eq:per}
	\end{equation}
	
	\item Dirichlet-Dirichlet (DD) boundary conditions
	\begin{equation}
		u_{L}^{({\rm DD})}(r,k) =\left[2/(L+1)\right]^{1/2}\sin r\varphi_{L}^{({\rm DD})}(k).
		\label{eq:DD}
	\end{equation}
	
	\item Neumann - Neumann (NN) boundary conditions
	\begin{equation}
		u_{L}^{({\rm NN})}(r,k) =\left \{ \begin{array}{ll} L^{-1/2} & \mbox{for} \quad k=1 \\ \left[2/L\right]^{1/2}\cos (r-1/2)\varphi_{L}^{({\rm NN})}(k) & \mbox{for} \quad  k=2,\cdots, L \end{array}\right..
		\label{eq:NN}
	\end{equation}
	
	\item Neumann - Dirichlet (ND) boundary conditions
	\begin{equation}
		u_{L}^{({\rm ND})}(r,k) =2(2L+1)^{-1/2}\cos (r-1/2)\varphi_{L}^{({\rm ND})}(k).
		\label{eq:DN}
	\end{equation}
	
\end{itemize}

The quantities $\varphi_L^{(\tau)}$, $k=1,\dots ,L$, are defined as follows 
\begin{equation} 
	\begin{array}{lll}
		&\varphi_{L}^{(p)}(k) = 2\pi k/ L, \quad   &\varphi_{L}^{\rm (DD)}(k) = \pi k/ (L+1), \\
		&\varphi_{L}^{\rm (NN)}(k) = \pi (k-1)/L, \quad 
		&\varphi_{L}^{({\rm ND})}(k) = \pi (2k-1)/(2L+1). 
	\end{array}
	\label{eq:angles}
\end{equation}

Now we are ready to find the finite-size behavior of the Gaussian model under periodic, Dirichlet-Dirichlet, Neumann-Neumann, and Neumann-Dirichlet boundary conditions. According to \eq{eq:per}  $S(0)=S(L)$, which corresponds to periodic boundary conditions; \eq{eq:DN} leads to $S(0)=S(1)$, i.e., one has there realization of Neumann boundary conditions, while $S(L+1)=0$, which corresponds to Dirichlet boundary conditions; finally \eq{eq:DD} leads to $S(0)=S(L+1)=0$, which is the realization of Dirichlet-Dirichlet boundary conditions. Thus, in the envisaged one-dimensional chain one has $L$ independent spin variables $\left \{S(1), S(2), \cdots, S(L)\right \}$.  

We start with the consideration of $d=3$ dimensional system with a film geometry. Note that:

\begin{itemize}
	\item under fully periodic (p) boundary conditions, $\tau =(p,p,p)$, one has
	${\mathbf k}_{0}=(L_1, L_2, L_3)$, hence $\mu_{\Lambda}^{(p, p ,p)}
	({\mathbf k}_{0})=6$. 
	
	\item under Dirichlet-Dirichlet boundary conditions along $z$ direction, i.e., $\tau =(p,p,{\rm DD})$, one has
	${\mathbf k}_{0}=(L_1, L_2, 1)$, with $\mu_{\Lambda}^{(p,p,{\rm DD})}
	({\mathbf k}_{0})=4+2\cos[\pi/(L+1)]$.
	
	\item under Neumann-Neumann boundary conditions along $z$ direction, i.e., $\tau =(p,p,{\rm NN})$, one has
	${\mathbf k}_{0}=(L_1, L_2, 1)$, with $\mu_{\Lambda}^{(p,p,{\rm NN})}
	({\mathbf k}_{0})=6$.
	
	\item under Neumann-Dirichlet boundary conditions along $z$ direction, i.e., $\tau =(p,p,{\rm ND})$, one has
	${\mathbf k}_{0}=(L_1, L_2, 1)$, hence $\mu_{\Lambda}^{(p,p,{\rm ND})}
	({\mathbf k}_{0})=4+2\cos[\pi/(2L+1)]$.
\end{itemize}

We set ${ \tau}=(p,p,p)$, ${ \tau}=(p,p,{\rm ND})$,  and ${ \tau}=(p,p,{\rm DD})$ for periodic, Neumann-Dirichlet and Dirichlet-Dirichlet boundary conditions, correspondingly, and use the short-hand notation $\tau=p$, $\tau={\rm ND}$ and $\tau={\rm DD}$ for them. Then, in order to arrive at a film geometry, we perform in \eq{eq:interaction-term-finite-region} the limits $L_1,L_2 \to \infty$, keeping $L_3=L$ fixed. 	For the interaction term one then obtains

\begin{itemize}
	\item Periodic boundary conditions
	\begin{equation}
		{\mathcal U}_{L,3}^{(p)}(K)= \lim_{L_{1},L_{2} \rightarrow \infty}
		{\mathcal U}_{\Lambda}^{(p,p,p)}(K) =  {1\over L}
		\sum_{k=1}^{L}{\mathcal V}_{2}\left[ 6 (K_c/K-1) + 2\left(1-
		\cos 2\pi\frac{k}{L}\right)\right];
		\label{eq:U-term-per}
	\end{equation}
	
	\item {Dirichlet--Dirichlet boundary conditions}
	\begin{equation}
		{\mathcal U}_{L,3}^{({\rm DD})}(K)= \lim_{L_{1},L_{2} \rightarrow \infty}
		{\mathcal U}_{\Lambda}^{(p,p,{\rm DD})}(K) =  {1\over L}
		\sum_{k=1}^{L}{\mathcal V}_{2}\left[ 6 (K_c/K-1) + 2\left(1-
		\cos \pi\frac{k}{L+1}\right)\right];
		\label{eq:U-term-DD}
	\end{equation}
	\item {Neumann--Neumann boundary conditions}
	\begin{equation}
		{\mathcal U}_{L,3}^{({\rm NN)}}(K)= \lim_{L_{1},L_{2} \rightarrow \infty}
		{\mathcal U}_{\Lambda}^{(p,p,{\rm NN})}(K) =  {1\over L}
		\sum_{k=1}^{L}{\mathcal V}_{2}\left[ 6 (K_c/K-1) + 2\left(1-
		\cos \pi\frac{k-1}{L}\right)\right];
		\label{eq:U-term-NN}
	\end{equation}
	and	
	\item {Neumann--Dirichlet boundary conditions}
	\begin{equation}
		{\mathcal U}_{L,3}^{({\rm ND})}(K)= \lim_{L_{1},L_{2} \rightarrow \infty}
		{\mathcal U}_{\Lambda}^{(p,p,{\rm ND})}(K) =  {1\over L}
		\sum_{k=1}^{L}{\mathcal V}_{2}\left[ 6 (K_c/K-1) + 2\left(1-
		\cos \pi\frac{2k-1}{2L+1}\right)\right].
		\label{eq:U-term-ND}
	\end{equation}
\end{itemize}

The calculations for Dirichlet -- Dirichlet boundary conditions are given in  \ref{sec:DD},  in \ref{sec:ND} for Neumann -- Dirichlet, in \ref{sec:NN} for Neumann -- Neumann, and periodic \ref{sec:periodic}.  

\subsection{Results for excess free energy, Casimir force, Helmholtz force and susceptibility} \label{sec:results1}

The scaling functions of the excess Gibbs free energy and the Casimir force for \textit{zero} field in the case of all boundary conditions listed above are  well known; see, e.g., Refs. \cite{BDT2000,DD2022}.  Here we will extend them to the case of nonzero field. Furthermore, we will present new results for the behavior of the fluctuation induced force in the canonical ensemble, i.e., for the Helmholtz force. The details of the calculations are organized in Appendices. Even when we reproduce there the results for the zero field, we will do that in a new way, specific to lattice systems. It is our hope that the technique can be used also in attacking other problems. In the main text we present the final results for the quantities involved and display plots of these under the corresponding boundary conditions.

\subsubsection{Dirichlet -- Dirichlet boundary conditions} 
\label{sec:Dirichlet-Dirichlet-bc}

The details of the derivation of the interaction and field terms are given in \ref{sec:DD}. The bulk of those details are shown below in the scaling regime 
\begin{equation}
	\label{eq:scaling-regime-DD}
	x_t= 6 (K_c/K-1) (L+1)^2 ={\cal O}(1), \quad \mbox{and} \quad 
		x_h=\beta^{-1/2}(L+1)^{3/2}L\; h. 
	\end{equation}
	
For the excess free energy one has 
\begin{equation}
	\label{eq:excess-free-energy-scaling-DD}
	\beta \Delta f_{{\rm ex},GC}^{({\rm DD})}(\beta,h=0)+\beta \Delta f_{{\rm ex},3}^{({\rm DD})}(\beta,h)=\frac{1}{L^2} \left[X_{\rm ex, \rm GC}^{(\rm DD)}\left(a_\beta t L^{1/\nu},h=0\right)+\Delta X_{\rm ex, \; GC}^{(\rm DD)}\left(a_\beta t L^{1/\nu},a_h h^{\Delta/\nu}|\; h\ne 0 \right)\right], 
\end{equation}
where $a_\beta$, $a_h$ are non-universal constants, and $X_{\rm ex}$ are universal scaling functions, $t=(T-T_c)/T_c$. For the excess free energy related to the field term, see \eq{eq:free-energy-density-finite-region}, one derives
\begin{equation}
	\label{eq:excess-field-term-DD}
	\beta \Delta f_{{\rm ex},3}^{({\rm DD})}(\beta,h)= -\frac{1}{2} L \left[P_{L}^{({\rm DD})}(h;\beta) - P_{\infty}(h;\beta) \right]=\frac{1}{2}\frac{x_h^2}{L^2} \frac{\tanh \left(\sqrt{x_t}/2\right)}{ x_t^{3/2}}.
\end{equation}
For the total excess free energy scaling function one has 
\begin{equation}
	\label{eq:scaling-function-excess-free-energy}
	X_{\rm ex\; GC}^{(\rm DD)}(x_t,x_h)=-\frac{1}{16 \pi} \left[2\sqrt{x_t}\; \text{Li}_2\left(e^{-2\sqrt{x_t}}\right)+\text{Li}_3\left(e^{-2\sqrt{x_t}}\right)\right]+\frac{1}{2}
	x_h^2 \frac{\tanh \left(\sqrt{x_t}/2\right)}{x_t^{3/2}}
\end{equation}
	 with
	\begin{equation}
		\label{eq:Casimir-amplitude-periodi}
		 \Delta_{\rm Casimir}^{(\rm DD))}\equiv X_{\rm ex,\;GC}^{(\rm DD)}(x_t=0,x_h=0)=-\frac{\zeta (3)}{16 \pi }. 
	\end{equation}

The corresponding result for the scaling function of the Casimir force is
\begin{equation}
	\label{eq:scaling-function-Casimir-force}
	X_{\rm Casimir}^{(\rm DD)}(x_t,x_h)=-\frac{1}{8 \pi} \left[2\sqrt{x_t}\; \text{Li}_2\left(e^{-2\sqrt{x_t}}\right)+\text{Li}_3\left(e^{-2\sqrt{x_t}}\right)-2 x_t \ln\left(1-e^{-2\sqrt{x_t}}\right)\right]-\frac{x_h^2 \;\text{sech}^2\left(\sqrt{x_t}/2\right)}{4 x_t}<0. 
\end{equation}
\begin{figure}[h!]
	\centering
	\includegraphics[width=3.in]{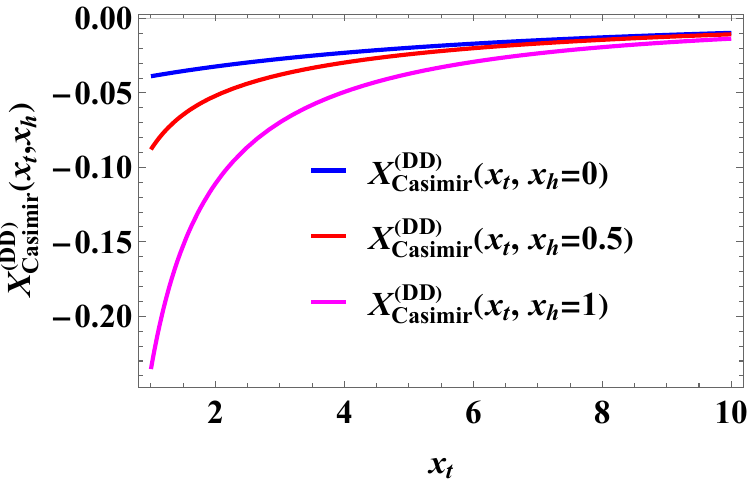} 
	\includegraphics[width=3.4in]{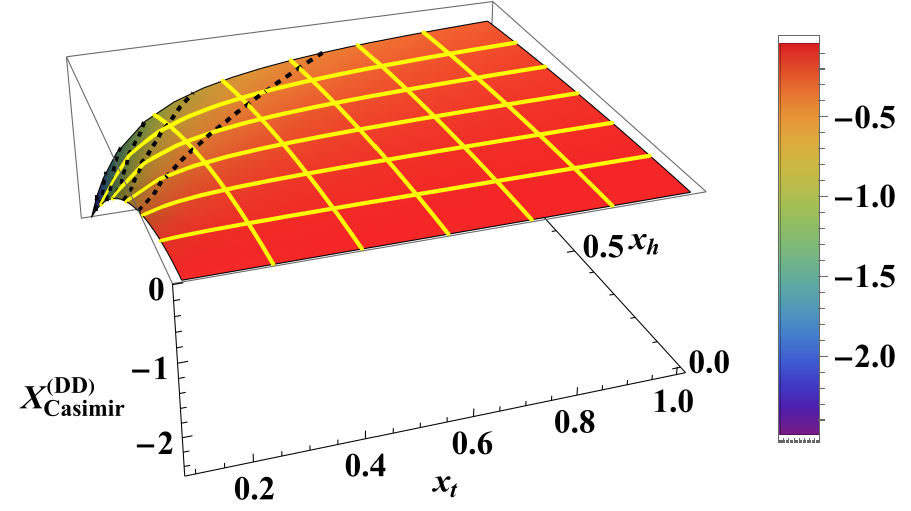} 
	\caption{The behavior of the  scaling function of the Casimir force $X^{\rm (DD)}_{(\rm Casimir)}(x_t,x_h)$ under Dirichlet-Dirichlet boundary conditions.  The force is always \textit{attractive}. It is strongest for nonzero field near $T=T_c$. }
	\label{fig:sscaling-function-Casimir-DD-zero-field}
\end{figure}

\begin{figure}[h!]
	\centering
	\includegraphics[width=5.2in]{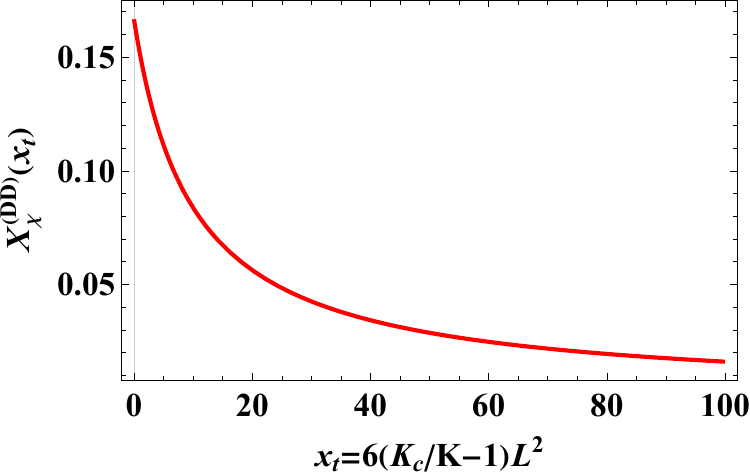} 
	\caption{The behavior of the  scaling function $X_\chi^{(\rm DD)}$. }
	\label{fig:scaling-function-Xy}
\end{figure}

For the behavior of the susceptibility in the finite system we derive the scaling behavior of the susceptibility under Dirichlet-Dirichlet boundary conditions
\begin{equation}
	\label{eq:scaling-susceptibility-DD}
	\beta \chi_{L}^{({\rm DD})}(\beta,h)=L^2 X^{(\rm DD)}_\chi(x_t), \quad \mbox{where} \quad  X^{(\rm DD)}_\chi(x_t) =
	\frac{2}{ x_t}\left[1-\frac{\tanh \left(\sqrt{x_t}/2\right)}{\sqrt{x_t}/2}\right]. 
\end{equation}
The behavior of the scaling function $X\chi^{(\rm DD)}(x_t)$ is given in Fig. \ref{fig:scaling-function-Xy}.

\begin{figure}[h!]
	\centering
	\includegraphics[width=5.2in]{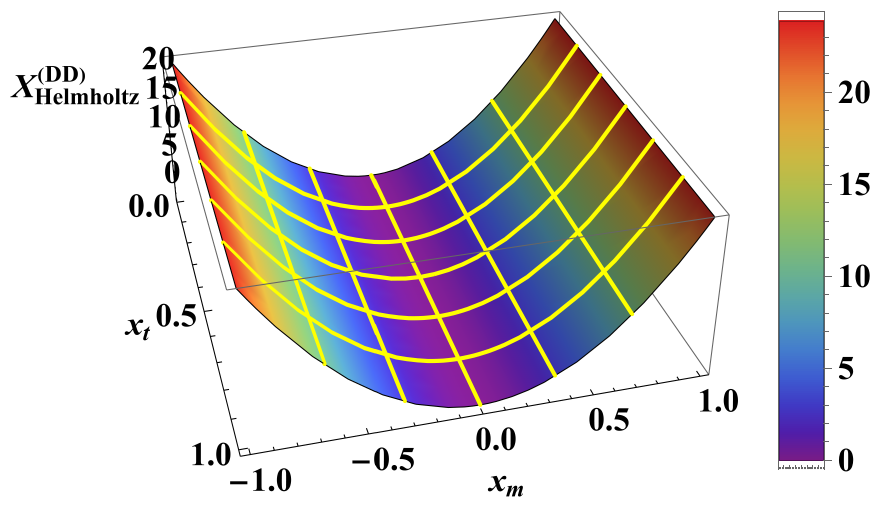} 
	\caption{The behavior of Helmholtz force as a function of $x_m$ and $x_t$. We observe that, depending on the  values of the scaling variable $x_t$ and $x_m$, it can be both attractive and repulsive.}
	\label{fig:Helmholtz-DD}
\end{figure}

As shown in \ref{sec:DD}  the scaling form of the $m$-dependent term in the Helmholtz free energy density of the canonical ensemble reads 
\begin{equation}
	\label{eq:Q-ND-final}
	Q_{L}^{\rm (DD)}(K,m;s)
	= {(L_3+1)}^{-3}2 x_t x_m^2 \left[1-\frac{\tanh \left(\sqrt{x_t}/2\right)}{\sqrt{x_t}/2}\right]^{-1},
\end{equation}
where
	\begin{equation}
	\label{eq:scaling-regime}
	x_t= 6 (\beta_c/\beta-1) (L+1)^2 ={\cal O}(1) \quad \mbox{and} \quad x_m= m \sqrt{K L}.
\end{equation}
As we see, this term also perfectly fits in a scaling form. From here we derive for the Helmholtz force 
\begin{eqnarray}
	\label{eq:helmholtz-DD}
	X_{\rm Helmholtz}^{(\rm DD)}(x_t,x_m) &=& -\frac{1}{8 \pi} \left[2\sqrt{x_t}\; \text{Li}_2\left(e^{-2\sqrt{x_t}}\right)+\text{Li}_3\left(e^{-2\sqrt{x_t}}\right)-2 x_t \ln\left(1-e^{-2\sqrt{x_t}}\right)\right] \\
	&& -\frac{4 x_m^2 x_t \left(x_t-2 \cosh \left(\sqrt{x_t}\right)+2\right)}{x_t-4 \sqrt{x_t} \sinh \left(\sqrt{x_t}\right)+(x_t+4) \cosh \left(\sqrt{x_t}\right)-4}. \nonumber
\end{eqnarray} 
Its behavior is visualized in Figure \ref{fig:Helmholtz-DD}.

\subsubsection{Neumann -- Dirichlet boundary conditions} 
\label{sec:Neumann-Dirichlet-bc}

The case of Neumann -- Dirichlet boundary conditions within the grand canonical ensemble has been already considered in Ref. \cite{Dantchev2025}. That is why for the $(T-h)$-ensemble we will juts state the corresponding results.

We identified there that 
\begin{equation}
	\label{eq:scaling-function-excess-free-energy}
	X_{\rm ex}^{(\rm ND)}(x_t,x_h)=-\frac{1}{16 \pi} \left[\sqrt{x_t}\; \text{Li}_2\left(-e^{-\sqrt{x_t}}\right)+\text{Li}_3\left(-e^{-\sqrt{x_t}}\right)\right]+x_h^2 \frac{\tanh \left(\sqrt{x_t}/2\right)}{2 x_t^{3/2}},
\end{equation}
where 
\begin{equation}
	\label{eq:scaling-regime-ND}
		x_h=\beta^{-1/2}(2L+1)^{3/2}L\; h, \quad x_t= 6 (\beta_c/\beta-1) (2L+1)^2,
\end{equation}
and that
\begin{equation}
	\label{eq:scaling-function-of-Casimir-zero-field}
	X_{{\rm Casimir}}^{(\rm ND)}(x_t, x_h)=-\frac{1}{8 \pi}\left\{\text{Li}_ 3\left(-e^{-\sqrt{x_t}}\right)+\sqrt{x_t} \text{Li}_ 2\left(-e^{-\sqrt{x_t}}\right)-\frac{1}{2} x_t \log \left(e^{-\sqrt{x_t}}+1\right)\right\}-\frac{x_h^2}{4}\left[\frac{\text{sech}^2\left(\sqrt{x_t}/2\right)}{x_t}\right].
\end{equation}
For the Casimir amplitude we have 
\begin{equation}
	\label{eq:Cas-amplitude}
	\Delta_{{\rm Casimir}}^{({\rm ND})}\equiv \frac{1}{2}X_{{\rm Casimir}}(x_t=0,h=0)=\frac{3 }{64 \pi }\zeta (3)>0. 
\end{equation}
\begin{figure}[h!]
	\centering
	\includegraphics[width=3.2in]{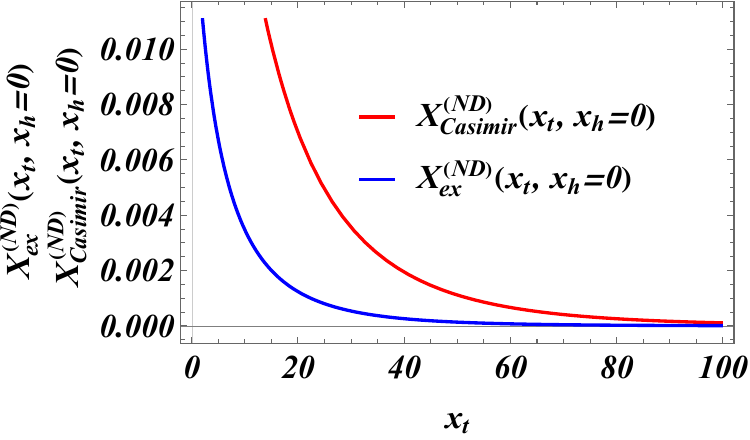} 
	\includegraphics[width=3.2in]{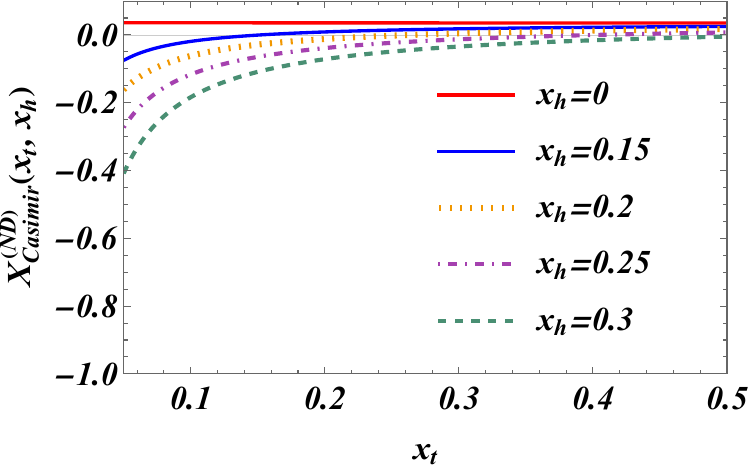} 
	\caption{The behavior of the excess free energy and Casimir force under Neumann -- Dirichlet boundary conditions as a function of the  temperature scaling variable $x_t$ (left panel).  For several fixed values of the field scaling variable $x_h$ the behavior of the forces is also visualized (the right panel). We observe that for zero field the force is always \textit{repulsive}.  However, when $h\ne 0$ the force becomes attractive very close to $T_c$ and slightly repulsive for moderate values of $x_h$ and large values of $x_t$.  For large values of $x_h$ the force is always attractive. }
	\label{fig:Casimir-ND}
\end{figure}

In \ref{sec:ND} we have shown that the therm in the Helmholtz free energy which depends on scaling variable $x_m$, related to the magnetization, is 
\begin{equation}
	\label{eq:Q-ND-final}
	Q_{L}^{\rm (ND)}(K,m;s)
	= {(2L_3+1)}^{-3} 2 x_t x_m^2 \left[1-\frac{\tanh \left(\sqrt{x_t}/2\right)}{\sqrt{x_t}/2}\right]^{-1}.
\end{equation}
where
\begin{equation}
	\label{eq:scaling-regime}
	x_t= 6 (\beta_c/\beta-1) (2L+1)^2 ={\cal O}(1) \quad \mbox{and} \quad x_m= m \sqrt{K L}.
\end{equation}
We see that, up to the non-universal prefactors in the definition of the scaling variables, this expression coinsides with the one for the Dirichlet -- Dirichlet boundary conditions - see \eq{eq:Q-ND-final}.

\begin{figure}[h!]
	\centering
	\includegraphics[width=5.2in]{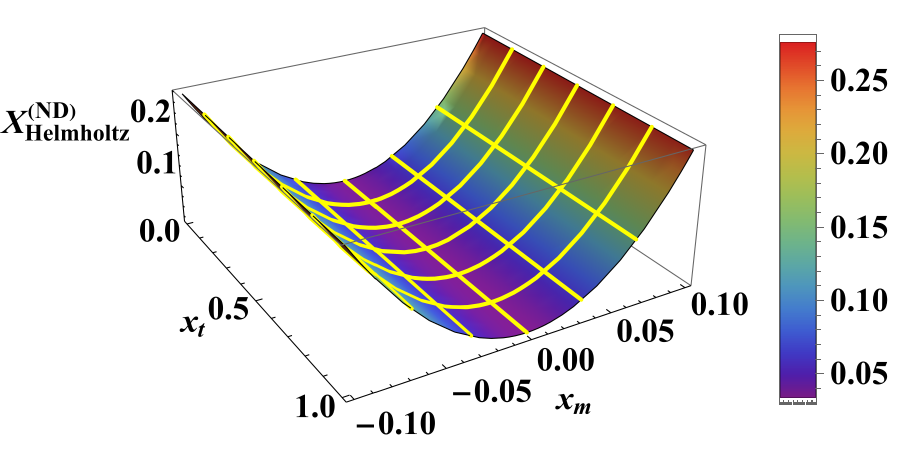} 
	\caption{The behavior of Helmholtz force as a function of $x_m$ and $x_t$. We observe that the force is always \textit{repulsive}.}
	\label{fig:Helmholtz-ND}
\end{figure}

As we see, this term also perfectly fits in a scaling form. From here we derive for the Helmholtz force 
\begin{eqnarray}
	\label{eq:helmholtz-DD}
	X_{\rm Helmholtz}^{(\rm DD)}(x_t,x_m) &=& -\frac{1}{8 \pi} \left[\sqrt{x_t}\; \text{Li}_2\left(-e^{-\sqrt{x_t}}\right)+\text{Li}_3\left(-e^{-\sqrt{x_t}}\right)-\frac{1}{2} x_t \ln\left(1+e^{-2\sqrt{x_t}}\right)\right] \\
	&& -\frac{4 x_m^2 x_t \left(x_t-2 \cosh \left(\sqrt{x_t}\right)+2\right)}{x_t-4 \sqrt{x_t} \sinh \left(\sqrt{x_t}\right)+(x_t+4) \cosh \left(\sqrt{x_t}\right)-4}. \nonumber
\end{eqnarray} 
Its behavior is visualized in Figure \ref{fig:Helmholtz-ND}.

\subsubsection{Neumann -- Neumann boundary conditions} 
\label{sec:Neumann-Neumann-bc}

As shown in \ref{sec:NN}, for the excess free energy under Neumann -- Neumann boundary conditions one obtains
\begin{equation}
	\label{eq:scaling-function-excess-free-energy-A}
	X_{\rm ex}^{(\rm NN)}(x_t)=-\frac{1}{16 \pi} \left[2\sqrt{x_t}\; \text{Li}_2\left(-e^{-2\sqrt{x_t}}\right)+\text{Li}_3\left(-e^{-2\sqrt{x_t}}\right)\right]. 
\end{equation}
From here one derives, similar to the Dirichlet -- Dirichlet boundary conditions case
\begin{equation}
	\label{eq:scaling-function-Casimir-force}
	X_{\rm Casimir}^{(\rm NN)}(x_t,x_h)=-\frac{1}{8 \pi} \left[2\sqrt{x_t}\; \text{Li}_2\left(e^{-2\sqrt{x_t}}\right)+\text{Li}_3\left(e^{-2\sqrt{x_t}}\right)-2 x_t \ln\left(1-e^{-2\sqrt{x_t}}\right)\right]<0. 
\end{equation}
Note that these expressions coincide with the ones resulting from       Dirichlet -- Dirichlet boundary conditions case with $h=0$.  The results for $h\ne 0$, however, differ. For Neumann -- Neumann boundary conditions nether the excess free energy, nor the Casimir force depend on $h$, which is \textit{not} the case for Dirichlet - Dirichlet boundary conditions. 

\begin{figure}[h!]
	\centering
	\includegraphics[width=5.2in]{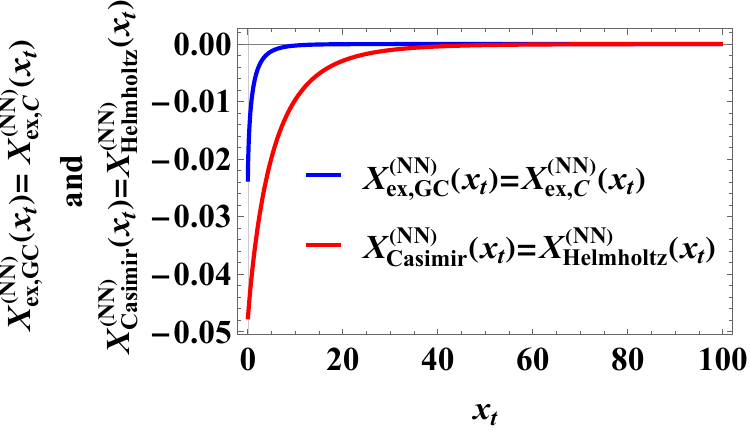} 
	\caption{The behavior of the \textit{i)} excess free energy in grand canonical and canonical ensembles under Neumann- Neumann boundary conditions. \textit{ii)} The bahvior of the Casimir and the Helmholtz force as a function  $x_t$. There is no dependence of the neither the field scaling variable $x_h$, nor one the magnetization one $x_m$. We observe that these forces are equal to each other and are always attractive. }
	\label{fig:Helmholtz-NN}
\end{figure}

\subsubsection{Periodic boundary conditions} 
\label{sec:per-boundary-conditions}

\begin{figure}[h!]
	\centering
	\includegraphics[width=5.2in]{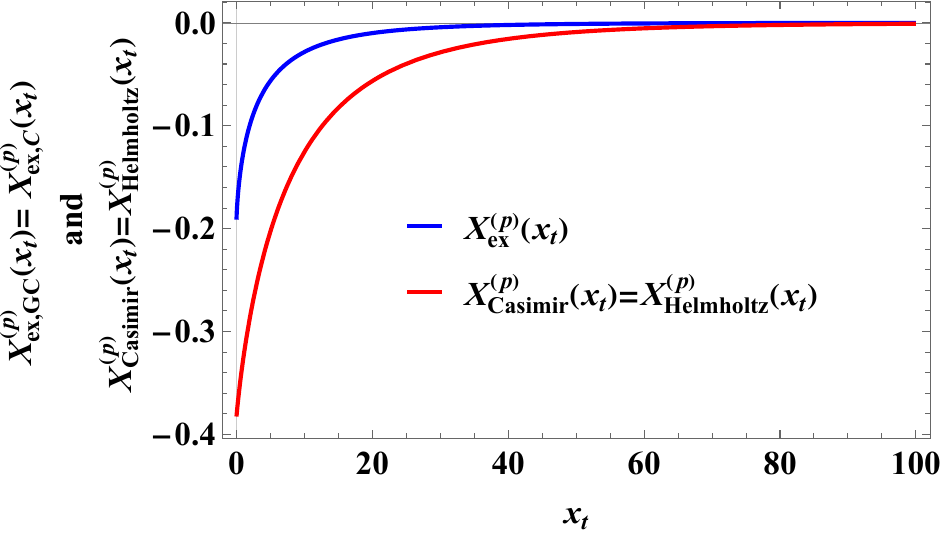} 
	\caption{The behavior of the  scaling function of the excess free energy  $X^{(\rm p)}_{\rm ex}(x_t)$ and the Casimir force $X^{(p)}_{\rm Casimir}(x_t)$ under periodic boundary conditions. These scaling functions equal those for a system in canonical ensemble, i.e. $X^{(p)}_{\rm Casimir}(x_t)= X^{(p)}_{\rm Helmholtz}(x_t)$. Here $x_t=6(K_c/K-1)L^2$.  The functions do \textit{not} depend on $h$ in the grand canonical ensemble, nor on $m$ in the canonical one.}
	\label{fig:sscaling-function-Casimir-periodic-zero-field}
\end{figure}

As  is demonstrated in \ref{sec:periodic}, for the excess free energy and the Casimir force we derive 
\begin{equation}
	\label{eq:scaling-function-excess-free-energy}
	X_{\rm ex, \; GC}^{(p)}(x_t)\equiv 	X_{\rm ex, \; GC}^{(p)}(x_t,h=0)= X_{\rm ex, \; GC}^{(p)}(x_t,h)=-\frac{1}{2 \pi} \left[\sqrt{x_t}\; \text{Li}_2\left(e^{-\sqrt{x_t}}\right)+\text{Li}_3\left(e^{-\sqrt{x_t}}\right)\right]<0,
\end{equation}
for the scaling function of the excess free energy in the grand canonical ensemble and 
\begin{equation}
	\label{eq:Casimir-periodic-zero-field}
	X_{\rm Casimir}^{(p)}(x_t,h)=	-\frac{1}{\pi}\left[\text{Li}_3\left(e^{-\sqrt{x_t}}\right)+\sqrt{x_t} \text{Li}_2\left(e^{-\sqrt{x_t}}\right)-\frac{1}{2} x_t \log \left(1-e^{-\sqrt{\text{xt}}}\right)\right],
\end{equation}
for the scaling function of the Casimir force.  Note that the both scaling functions do \textit{not} depend on $h$. Their behavior on the scaling variable 
\begin{equation}
	\label{eq:scaling-regime-A}
	x_t= 6 (K_c/K-1) L^2
\end{equation}
is shown in Fig. \ref{fig:sscaling-function-Casimir-periodic-zero-field}. 
It is easy to show that when $x_t\to 0$
\begin{equation}
	\label{eq:Cas-amplitude-periodic}
	\Delta^{(p)}_{\rm Casimir}\equiv X_{\rm ex}^{(p)}(x_t=0)=	-\frac{\zeta (3)}{2\pi }. 
\end{equation}
The scaling functions of the excess Gibbs free energy and the Casimir force for zero field are  well known - see, e.g., Refs. \cite{BDT2000,DD2022}. The new information here is, that the field term for the
finite system is equal, as shown in \ref{sec:periodic}, to the one of the bulk system.  Thus, there is no  field dependence in the excess free energy and Casimir force.  For any field the corresponding functions are equal to  the ones for zero field. 

The dependence on $M$ in the canonical ensemble is given by $Q_{\Lambda}^{\rm (per)}(K,M;s)$. For the case of periodic boundary conditions this term is easily evaluated in any $d$. One has
\begin{eqnarray}
	\label{eq:Q-periodic}
	Q_{\Lambda}^{\rm (per)}(K,M;s)&=&\left[ \frac{1}{|\Lambda|} \ln \frac{2\pi}{K} {|\Lambda| \over{ 2s/K - \mu_{\Lambda}^{\rm (per)}({\mathbf 0})}}\right]+  \frac{ K M^2}{|\Lambda|^2} \left[2s/K - \mu_{\Lambda}^{\rm (per)}({\mathbf 0})\right] \nonumber \\
	&=& \left[ \frac{1}{|\Lambda|} \ln \frac{2\pi}{K} {|\Lambda| \over{ 2d (K_c/K-1)}}\right] +  \frac{ K M^2}{|\Lambda|^2} \left[2d (K_c/K-1) \right],
\end{eqnarray}
where we have taken into account that $\mu_{\Lambda}^{(p,p,p)}({\mathbf 0})=2d$. Here we have used the short-hand notation $\tau={\rm (per)}$ for periodic boundary conditions. Since we are mainly interested in the behavior of the system with a film geometry, we perform in \eq{eq:Q-periodic} the limits $L_1,L_2 \to \infty$, keeping $L_3=L$ fixed. Introducing the $m^2\equiv \lim_{|\Lambda| \to \infty} M^2/|\Lambda|^2$, we obtain 
\begin{equation}
	\label{eq:Q-periodic-final}
	Q_{L}^{\rm (per)}(K,m;s)
	= K m^2 \left[2d (K_c/K-1) \right].
\end{equation}
Obviously, this term does not depend on $L$ and, therefore, it equals the corresponding term in the \textit{bulk} system, i.e.,
\begin{equation}
	\label{eq:bulk-term-m-Helmhotz}
	Q_{\rm bulk}(K,m;s)
	= K m^2 \left[2d (K_c/K-1) \right].
\end{equation}
Thus, from \eq{eq:free-energy-density-finite-region-M} and \eq{eq:Q-periodic-final} it follows that the Helmholtz force in the Gaussian model under periodic boundary conditions does \textit{not} depend on $m$ and is equal to the Helmholtz force of this model in the grand canonical ensemble with  zero external field. Therefore, it is a force of attraction. It can easily be shown that the above results for periodic boundary conditions is valid for $2<d<4$. It is easy to see that, for such $d$, it can be written in a scaling form
\begin{eqnarray}
	\label{eq:Q-periodic-final-scaling-form}
	Q_{L}^{\rm (per)}(K,m;s)= L^{-d} \; x_m^2 \;x_t,
\end{eqnarray} 
where 
\begin{equation}
	\label{eq:scaling-variables-m}
	x_t=\left[2d (K_c/K-1) \right]\; L^{1/\nu} , \quad \mbox{with} \quad \nu=1/2, \quad \mbox{and} \quad  x_m = \sqrt{K}\; m \; L^{\beta/\nu}, \quad \mbox{with} \quad \beta =(d-2)/4. 
\end{equation}

For the susceptibility under periodic boundary conditions, from \eq{eq:P-per-scaling} one derives 
\begin{equation}
	\label{eq:susceptibility-per}
	\beta 	\chi_{L}^{({\rm per})}(\beta,h) = L^2 \;\frac{2}{ x_t}=  L^2 \;X^{(\rm per)}_\chi( x_t), \quad \mbox{with} \quad X^{(\rm per)}_\chi( x_t)=\frac{2}{ x_t}. 
\end{equation}
\begin{figure}[h!]
	\centering
	\includegraphics[width=5.2in]{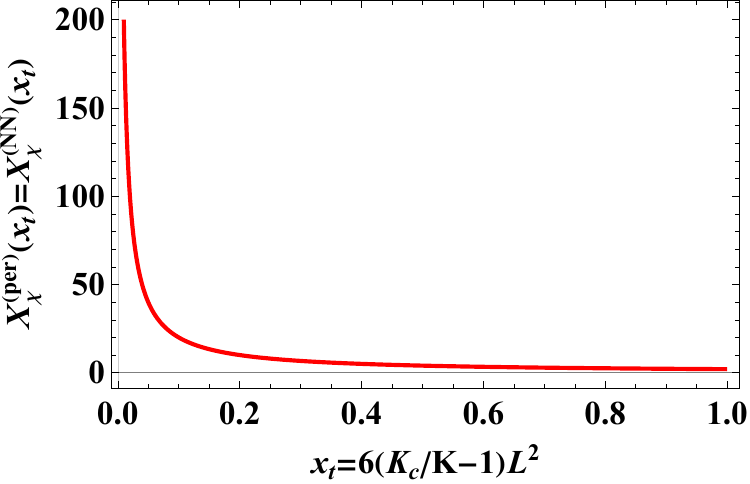} 
	\caption{The behavior of the  scaling function of the susceptibility $X^{(\rm per)}_\chi(x_t)$ under periodic boundary conditions.Here $x_t=6(K_c/K-1)L^2$. Note that the scaling function does \textit{not} depend on $x_h$. As we will see below the same scaling function provides also the behavior for the Neumann-Neumann boundary conditions. }
	\label{fig:scaling-function-susceptibility-periodic}
\end{figure}

Thus, Fig. \ref{fig:sscaling-function-Casimir-periodic-zero-field} shows, as indicated there, \textit{both} the Casimir force, \textit{and} the Helmholtz force in the Gaussian model under periodic boundary conditions.As we have seen this is \textit{not} the case for Dirichlet -- Dirichlet boundary conditions. 

\section{The continuum version of the Gaussian model: preliminary comments and scaling variables}
\label{sec:continuum}

As an alternative approach leading to precisely the same results in the scaling regime as those obtained in Sections \ref{sec:per-boundary-conditions} to \ref{sec:Neumann-Dirichlet-bc}, we consider the properties discussed above of a \textit{ continuum }version of the Gaussian model. We will be interested in the behavior of  these properties in the scaling regime. To this end we define a set of scaling parameters \cite{G92,BDT2000}. They are
\begin{eqnarray}
x_t &=& tL^{1/\nu} \quad \rightarrow \quad x_t
=  tL^2,  \label{eq:prelima} \\
x_h&=&hL^{(d+2-\eta)/2} \quad \rightarrow \quad 
x_h= hL^{(d+2)/2}, \label{eq:prelimb} \\
x_m&=& hL^{(d-2+\eta)/2} \quad \rightarrow \quad 
x_m = hL^{(d-2)/2} \label{eq:prelimc}.
\end{eqnarray}
In the above, $t$ is the reduced temperature, $h$ is the ordering field and $m$ is the magnetization per unit volume. The quantity $d$ is the dimensionality, $\nu$ is the correlation length exponent and $\eta$ is the anomalous dimension. The second post in each equation above replaces $\nu$ and $\eta$ by the values that are appropriate to the Gaussian model. We take all non-universal amplitudes to be equal to 1. 

For future reference, we will be making use of the dimensional factor 
\begin{equation}
\mathcal{K}_d=\frac{2 \pi^{d/2}}{\Gamma \left( \frac{d}{2}\right)} \label{eq:prelim4}
\end{equation}
This factor arises when we integrate over wave-vectors lying in the plane occupied by the finite-thickness slab. 
 
 In the grand canonical ensemble there will be two contributions to the Casimir force. The first is independent of the ordering field, $h$ and the second---when it is non-zero---is quadratic in that amplitude. In all cases, the Casimir force can be expressed as a function of the scaling quantities $x_t$ and $x_h$ divided by $L^d$, where $L$ is the thickness of the slab for which the Casimir force is calculated. 
 
 Similar comments apply in the case of  the canonical ensemble, for which  the Helmholtz force can be expressed as a function of the scaling parameters $x_t$ and $x_m$, also divided by $L^d$.

\subsection{Formulation, general results and definitions}
The continuum version of the Gaussian model with a scalar order parameter consists of the linear and bilinear terms in the Ginzburg-Landau-Wilson formulation of a system in $d$ dimensions that undergoes a continuous symmetry-breaking phase transition at low temperatures. The grand partition function of this system is the functional integral
\begin{equation}
\mathcal{Z}_G(t,h) = \int \exp[-\mathcal{F}(\psi(\vec{r}))] \, \mathcal{D} \{\psi(\vec{r})\} \label{eq:prelim1}
\end{equation} 
where
\begin{equation}
\mathcal{F}(\psi(\vec{r})) = \int \left[t \psi(\vec{r})^2+| \vec{\nabla} \psi(\vec{r})|^2 -h \psi(\vec{r})\right] \, d^dr \label{eq:prelim2}
\end{equation}
In (\ref{eq:prelim2}) $t$ is the reduced temperature, proportional to $T-T_c$, and $h$ is the spatially constant ordering field. Because of the Gaussian  nature of the free energy functional $\mathcal{F}(\psi(\vec{r}))$ the partition function resolves into the product
\begin{equation}
\mathcal{Z}_G(t,h) = \mathcal{Z}_{G,I}(t) \times \mathcal{Z}_{G,h}(t,h) \label{eq:prelim3}
\end{equation}
where $\mathcal{Z}_{G,I}(t)$ is the partition function of the system with $h=0$. The geometry of the system under consideration is a slab of large---ultimately infinite---cross section, and finite thickness $L$.

Our end-results for the Casimir forces acting upon the systems will depend on the boundary conditions imposed. In all cases, the form of the Casimir force is
\begin{eqnarray}
f_{\rm{Cas}}(t,h,L) & = & L^{-d} \left(X_{\rm {Casimir}}(x_t) +x_h^2P_{\rm{Casimir}}(x_t)\right)  \label{eq:prelim5a}
\end{eqnarray}

All results reported here rely on two relations, which can be obtained with the use of  contour integration techniques; see  \cite{GR}. Those relations are
\begin{eqnarray}
\sum_{n=-\infty}^{\infty} \frac{1}{an^2+b} & = & \frac{\pi \coth(\pi\sqrt{b/a})}{\sqrt{ab}}  \label{eq:prelim6} \\
\sum_{n=0}^{\infty} \frac{1}{c(2n+1)^2 +d} & = & \frac{\pi \tanh(\pi /2\sqrt{d/c}) }{4\sqrt{cd}} \label{eq:prelim7}
\end{eqnarray}

\subsection{The Casimir and Helmholtz force  of the continuum Gaussian model with Dirichlet-Dirichlet boundary conditions} 

\label{sec:Dirichlet-Dirichlet}

As the first case, we consider Dirichlet-Dirichlet boundary conditions. In order to evaluate the partition function and grand partition function, we construct a set of orthonormal eigenfunctions of the free energy functional in (\ref{eq:prelim1}) and (\ref{eq:prelim2}). These eigenfunctions have the form
\begin{equation}
\psi_{DD}^{(n)}(z,L,\vec{Q},\vec{R}) = \frac{1}{\sqrt{2LA}}\sin(n \pi z/L) \exp(i \vec{Q} \cdot \vec{R}) \label{eq:DD1}
\end{equation}
Where $A$ is the cross-sectional $d-1$-dimensional area of the slab, to be taken to infinity. It is readily established that
\begin{equation}
\int_{0}^L  \int_{\vec{R}\, \in\, \rm{slab} } \left|  \psi_{DD}^{(n)}(z,L,\vec{Q}, \vec{R})\right|^2  dz\, d^{d-1}R =1 \label{eq:DD2}
\end{equation}
So the eigenfunctions are properly normalized. We can also confirm that they are mutually orthogonal. We can also confirm that
\begin{equation}
\int_{0}^L  \int_{\vec{R}\, \in\, \rm{slab} } \left[  \psi_{DD}^{(n)}(z,L,\vec{Q}, \vec{R})\right]  dz\, d^{d-1}R =\delta(\vec{Q}) \left\{ \begin{array}{ll} 2\sqrt{2LA}/n \pi & n \, \rm{odd} \\ 0 & n \, \rm{even} \end{array} \right.  \label{eq:D3}
\end{equation}
\begin{figure}[htbp]
\begin{center}
\includegraphics[width=5in]{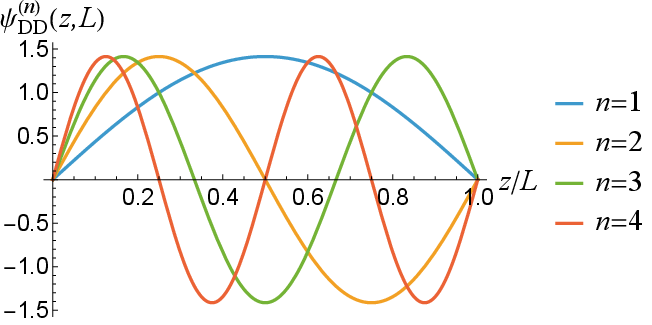}
\caption{The normalized eigenfunctions $\psi_{DD}^{(n)}(z,L)$, as defined in (\ref{eq:DD1})}
\label{fig:DDplot1}
\end{center}
\end{figure}

We start by expanding the order parameter of the system (\ref{eq:dnh5}), as
\begin{equation}
\Psi(z) = \sum_{n=1}^{\infty}a_n^{(DD)}\psi_{DN}^{(n)}(z,L,\vec{Q}, \vec{R}) \label{eq:DD4}
\end{equation}
Given this, the Gibs free energy functional in terms of the coefficients $a_n^{(DD)}$ is
\begin{equation}
G^{(DD)}(\{a_j\},t,A,L,h,\vec{Q}) = \sum_{n=1}^{\infty} (a_n^{(DD)})^2\left[\left(\frac{n \pi}{L} \right)^2 + t + Q^2 \right] - \sum_{m=0}^{\infty} h\frac{2\sqrt{2LA}}{2m+1} a_{2m+1}^{(DD)} \label{eq:DD5}
\end{equation}
We obtain the grand partition function by exponentiating the above expression and then performing the Gaussian integrals over the coefficients $a_j^{(DD)}$. This yields the product of two terms. They are as follows

\subsection{First term}
This is the result of the Gaussian integration if $h=0$. It is
\begin{equation}
T_1 = \exp \left[ \int\sum_{n=1}^{\infty} -\frac{1}{2} \ln \left(\left(\frac{n \pi}{L} \right)^2 + t + Q^2 \right) \frac{d^{d-1}Q}{(2 \pi)^{d-1}} \right] \label{eq:DD6}
\end{equation}
The next step in the full evaluation of this term is the summation over $n$. To carry this out, we take the derivative with respect to $t$ of the expression. This leaves us with the new exponent
\begin{equation}
- \frac{1}{2} \int \sum_{n=1}^{\infty} \frac{1}{\left(\frac{n \pi}{L} \right)^2 +t+Q^2} d^{d-1}Q = \int -\frac{L \sqrt{Q^2+t} \coth \left(L \sqrt{Q^2+t}\right)-1}{4 \left(Q^2+t\right)} \frac{d^{d-1}Q}{(2 \pi)^{d-1}} \label{eq:DD7}
\end{equation}
Integrating the integrand in (\ref{eq:DD7}) with respect to $t$, we are left with
\begin{eqnarray}
\lefteqn{\frac{1}{(2 \pi)^{d-1}}\int \frac{1}{4} \left(\log \left(Q^2+t\right)-2 \log \left(\sinh \left(L
   \sqrt{Q^2+t}\right)\right)\right) d^{d-1}Q} \nonumber \\ & = & \frac{1}{(2 \pi)^{d-1}}\int_0^{\infty} \frac{1}{4} \left(\log \left(Q^2+t\right)-2 \log \left(\sinh \left(L
   \sqrt{Q^2+t}\right)\right)\right) \mathcal{K}_{d-1} Q^{d-2}dQ  \label{eq:DD8}
\end{eqnarray}
To find the Casimir force we take the derivative with respect to $L$ of the result above. The result of this is 
\begin{equation}
-\frac{1}{2} \int_0^{\infty} \sqrt{Q^2+t}\coth(L \sqrt{Q^2+t}) \frac{\mathcal{K}_{d-1}}{(2 \pi)^{d-1}} Q^{d-2} dQ \label{eq:DD9}
\end{equation}
The Casimir force is the difference between the expression above and its limit as $L \rightarrow \infty$. The final result is
\begin{equation}
-\frac{1}{2} \int_0^{\infty} \sqrt{Q^2+t} \left(\coth(L\sqrt{Q^2+t}) -1 \right)\frac{\mathcal{K}_{d-1}}{(2 \pi)^{d-1}} Q^{d-2} dQ \label{eq:DD10}
\end{equation}
We can cast this into a scaling form by writing $Q=wL^{-1}$. Then if we introduce the scaling variable $x_t$ with
\begin{equation}
t=x_tL^{-2} \label{eq:DD11}
\end{equation}'
the expression in (\ref{eq:DD10})
becomes
\begin{equation}
-\frac{1}{2}L^{-d} \int_0^{\infty}\sqrt{w^2+x_t}\left(\coth\left(\sqrt{w^2+x_t}\right) -1 \right)\frac{\mathcal{K}_{d-1}}{(2 \pi)^{d-1}}w^{d-2} dw \label{eq:DD12}
\end{equation}
The scaling form of the Casimir force when $h=0$ 
\begin{equation}
X^{(DD)}_{\rm{Casimir}}(x_t) = L^d F^{(DD)}_{\rm{Casimir}}(x_t) \label{eq:DD13}
\end{equation}
is
\begin{equation}
X^{(DD)}_{\rm{Casimir}}(x_t) = -\frac{1}{2} \int_0^{\infty}\sqrt{w^2+x_t}\left(\coth\left(\sqrt{w^2+x_t}\right) -1 \right)\frac{\mathcal{K}_{d-1}}{(2 \pi)^{d-1}}w^{d-2} dw \label{eq:DD14}
\end{equation}
Here,  $K_n$  is the geometrical factor $2 \pi^{n/2} /\Gamma(n/2)$. 
Setting $d=3$ and making use of a change of variables, replacing $w^2$ with $W$, we end up with the following result, which holds when $d=3$.
\begin{equation}
X^{(DD)}_{\rm{Cas},1}(x_t) = -\frac{1}{4 \pi} \left(\sqrt{x_t} \text{Li}_2\left(e^{-2 \sqrt{x_t}}\right)+\frac{1}{2}
   \text{Li}_3\left(e^{-2 \sqrt{x_t}}\right)-x_t \log \left(1-e^{-2
   \sqrt{x_t}}\right)\right) \label{eq:DD15}
\end{equation}

\subsection{Second term}
If $h \neq 0$, there is an additional multiplicative term in the partition function. It is
\begin{eqnarray}
\lefteqn{\exp \left[\sum_{m=0}^{\infty} h^2 \frac{2LA}{(2m+1)^2 \pi^2} \frac{1}{((2m+1) \pi/L)^2+t}\right] } \nonumber \\ & = & \exp \left[ \frac{A h^2 \left(L \sqrt{t}-2 \tanh \left(\frac{L \sqrt{t}}{2}\right)\right)}{4 t^{3/2}} \right] \label{eq:DD16}
\end{eqnarray}
From this we infer an $h$-dependent contribution to the free energy per unit area given by
\begin{equation}
 -\frac{h^2 \left(L \sqrt{t}-2 \tanh \left(\frac{L \sqrt{t}}{2}\right)\right)}{4 t^{3/2}}. \label{eq:DD17}
\end{equation}
Taking the derivative with respect to $L$ of minus (\ref{eq:DD16}), with the limit of that expression for infinite $L$ subtracted, we end up with the result for the Casimir force per unit area
\begin{equation}
-\frac{h^2 \text{sech}^2\left(\frac{L \sqrt{t}}{2}\right)}{4 t} \label{eq:DD18}
\end{equation}
Making use of the scaling variables $x_t=tL^2$ and $x_h=h^{(d+2)/2}$ we end up with the result for the scaling form of the $h$-dependent Casimir force
\begin{equation}
X^{(DD)}_{\rm{Casimir}}(x_t, x_h) = - \frac{x_h^2 \mathop{\rm{sech}}^2(\sqrt{x_t}/2)}{4x_t} \label{eq:DD19}
\end{equation}

To calculate the Helmholtz force, we replace $h$ with $i \omega$ in (\ref{eq:DD16}) and insert $- i \omega M$ in the exponent, then integrate over $\omega$. The integral of interest is
\begin{equation}
\int_{-\infty}^{\infty} \exp \left[ \frac{- \omega^2 (L\sqrt{t}-2 \tanh(L\sqrt{t}/2))}{4t^{3/2}} + i \omega M\right] \label{eq:DD20}
\end{equation}
The term of interest in the result is the one going as $M^2$ in the exponent. This is
\begin{equation}
\exp \left[  \frac{-M^2 t^{3/2}}{L\sqrt{t} -2 \tanh(L\sqrt{t}/2)}\right] \label{eq:DD21}
\end{equation}
This implies a contribution to the free energy per unit area going as
\begin{equation}
\frac{M^2 t^{3/2}}{L\sqrt{t}- 2 \tanh(L\sqrt{t}/2)} = \frac{m^2Lt}{1-2 \tanh(L\sqrt{t}/2)/L\sqrt{t}} \label{eq:DD22}
\end{equation}
The excess free energy is the difference between the above expression and $m^2Lt$, the free energy per unit area in the bulk. Taking minus the $L$ derivative of that difference, multiplying by $L^d$, and replacing $m$ and $t$ by their associated scaling variables, we end up with the scaling contribution to the $m$-dependent  Helmholtz force for Dirichlet-Dirichlet boundary conditions
\begin{equation}
Q^{(DD)}_{{\rm Helmholtz}} = \frac{x_m^2 x_t \left(x_t-2 \cosh \left(\sqrt{x_t}\right)+2\right)}{\left(\sqrt{x_t}
   \cosh \left(\frac{\sqrt{x_t}}{2}\right)-2 \sinh
   \left(\frac{\sqrt{x_t}}{2}\right)\right){}^2} \label{eq:DD23}
\end{equation}
This expression is equivalent to the $m$-dependent contribution to  Eq. (\ref{eq:helmholtz-DD-A}).

The cases of Neumann-Dirichlet, periodic and Neumann-Neumann boundary conditions are addressed in   \ref{app:Neumann-Dirichlet}, \ref{app:periodic} and \ref{app:Neumann-Neumann}.

\section{Concluding remarks and discussion}
\label{sec:summary}

In the current article we investigate the ensemble dependence of two fluctuation induced forces---the Casimir force in the grand canonical ensemble and Helmholtz force in the canonical ensemble---in the case of one of the foundational models of the statistical mechanics, the Gaussian model. We consider Dirichlet-Dirichlet, Neumann-Dirichlet, Neumann-Neumann, and periodic boundary conditions.  We study  the behavior of the Casimir force as a function of $T$ and $h$, and of the Helmholtz force as a function of $T$ and $m$. 

Among the other results reported in the article are: 
\begin{itemize}
	
	\item For every boundary condition examined, we confirm that both forces follow finite-size scaling.

	\item For Dirichlet--Dirichlet and Neumann--Dirichlet boundary conditions the Casimir and the Helmholtz force differ from each other - see Figs. \ref{fig:sscaling-function-Casimir-DD-zero-field} and \ref{fig:Helmholtz-NN}.  For Dirichlet--Dirichlet boundary conditions the Casimir force is always attractive, while the Helmholtz force can be both attractive and repulsive as a function of $T$ and $m$. 
	
	\item For Neumann--Dirichlet boundary conditions the Casimir force changes sign from repulsive to attractive with increase of $h$, while the Helmholtz force stays always repulsive  --- see Fig. \ref{fig:Helmholtz-ND}  for the Helmholtz force	and Figs. (10) and (11) in Ref.\cite{Dantchev2025} for the Casimir force.  
	
	\item Under periodic and Neumann-Neumann boundary conditions the Casimir force and the Helmholtz force coincide - the first does not depend on $h$, while the latter does not depend on $m$; they are always attractive---see Fig. 		\ref{fig:sscaling-function-Casimir-periodic-zero-field}.

	\item In addition to the results obtained for the fluctuation induced forces we have derived, as a byproduct, exact results for the susceptibility under the various boundary conditions. These results are visualized in  Fig. \ref{fig:scaling-function-Xy} for Dirichlet--Dirichlet boundary conditions, Fig. \ref{fig:scaling-function-susceptibility-periodic} for periodic and Neumann-Neumann ones.  The corresponding result for Neumann--Dirichlet boundary conditions are presented in Ref. \cite[see Fig. 7]{Dantchev2025}. The corresponding analytical expression are given in \eq{eq:scaling-susceptibility-DD}, \eq{eq:susceptibility-per} for both Neumann--Neumann and periodic boundary conditions,  and Eq. (94) in \cite{Dantchev2025} for Neumann--Dirichlet,  receptively.  
\end{itemize}

It is worthwhile noting that the issue of ensemble dependence of the fluctuation induced force pertinent to a  given ensemble is by no means limited to the Gaussian model, or to the canonical ensemble. Calculations, similar to those what we have presented can, in principle, be performed for any ensemble, say, also for a micro-canonical ensemble,  and for any statistical mechanical model of interest, say, the spherical model for any $d$, or Potts, or spin-1 Ising chain, for which exact results seems feasible to obtain. Naturally, all that can be attacked numerically via, say, Monte Carlo techniques.

\appendix

\section{\textit{\textbf{The behavior of the interaction term, the field term, the excess free energy and the Casimir force under Dirichlet -- Dirichlet boundary conditions}}}
\label{sec:DD}

\subsection{Evaluation of the interaction term}

Explicitly, from \eq{eq:U-term-DD} one obtains
\begin{equation}
	\label{eq:uterm-explicit-DD}
	{\mathcal U}_{L,3}^{({\rm DD})}(\beta)=	 {1\over L}
	\sum_{k=1}^{L}{\mathcal V}_{2}\left[ 6 (K_c/K-1) + 2\left(1-
	\cos \pi\frac{k}{L+1}\right)\right]=	\frac{1}{(2\pi)^2}\, \int_{-\pi}^{\pi} {\mathrm d}\theta_1\int_{-\pi}^{\pi} {\mathrm d}\theta_2\;	S^{({\rm DD})}(\beta,L|\theta_1,\theta_2),
\end{equation}
with 
\begin{equation}
	\label{eq:S-DD-expl}
	S^{({\rm DD})}(\beta,L|\theta_1,\theta_2)=\frac{1}{L}			\sum_{k=1}^{L}\,{\ln}\left[ 6 (K_c/K-1) + 2\sum_{\nu =1}^2
	(1-\cos\theta_{\nu}) +2\left(1 -
	\cos \pi\frac{k}{L+1}\right)\right].	
\end{equation}

This sum is of the form
\begin{equation}
	\label{eq:definition-of-the-sum-DD}
	S^{({\rm DD})}(x,L)=\frac{1}{L}		\ln	\prod_{k=1}^{L}\,2\left[ \cosh(x) -
	\cos \pi\frac{k}{L+1}\right].
\end{equation}
The summations in \eq{eq:definition-of-the-sum-DD} can be performed using \cite{GR} the identity
\begin{equation}
	\label{eq:identity-DD}
	\prod_{k=1}^{L}\,2\left[ \cosh(x) -
	\cos \pi\frac{k}{L+1}\right]=\frac{\sinh\left[(L+1)x\right]}{\sinh(x)}. 
\end{equation}
With the help of the identity one derives
\begin{equation}
	\label{eq:the-sum-DD}
	S^{({\rm DD})}(x,L)= \frac{1}{L} \ln \frac{\sinh\left[(L+1)x\right]}{\sinh(x)}. 
\end{equation}
Obviously $	\lim_{L\to \infty} S^{({\rm DD})}(x,L)=x$.  Thus, the part of the excess free energy under Dirichlet -- Dirichlet boundary conditions that depends only on the interaction term is 
\begin{eqnarray}
	\label{eq:excess-DD-slab-interaction}
	\beta \Delta f_{{\rm ex},3}^{({\rm DD})}(\beta,h=0)&=& \frac{1}{2}\, L\, \left[{\mathcal U}_{L,3}^{({\rm ND})}(\beta)- 	{\mathcal U}_{\infty,3}(\beta) \right] \nonumber\\
	&=&  \frac{L}{8\pi^2}\, \int_{-\pi}^{\pi} {\mathrm d}\theta_1\int_{-\pi}^{\pi} {\mathrm d}\theta_2\;\left[ \frac{1}{L} \ln \frac{\sinh\left[(L+1)x\right]}{\sinh(x)} -x\right] \nonumber\\
	&=& \frac{1}{8\pi^2}\, \int_{-\pi}^{\pi} {\mathrm d}\theta_1\int_{-\pi}^{\pi} {\mathrm d}\theta_2\;\left[ \ln \left(\frac{1-e^{2(L+1) x}}{1-e^{-2 x}}\right) \right].
\end{eqnarray}

Let us consider the behavior of $\beta \Delta f_{{\rm ex},3}^{({\rm DD})}(\beta,h=0)$ in the scaling regime 
\begin{equation}
	\label{eq:scaling-regime-DD-A}
	x_t= 6 (K_c/K-1) (L+1)^2 ={\cal O}(1).
\end{equation}
Obviously, if $x={\cal O}(1)$ then the $L$-dependent part of $\beta \Delta f_{{\rm ex},3}^{({\rm DD})}(\beta,h=0)$ will be exponentially small which respect to $L$ and will also contain ${\cal O}(1)$ $L$-independent terms. Thus, we need to consider the regime $2(L+1)x={\cal O}(1)$. It follows that $x\ll 1$.  Using the same procedure as in \eq{eq:def-x} and \eq{eq:xl-r}, we arrive at 
\begin{eqnarray}
	\label{eq:g1-int-r}
	\beta \Delta f_{{\rm ex},3}^{({\rm DD})}(\beta,h=0) & \simeq & \frac{1}{8\pi} \int_{0}^{R}\left[ \ln \left(\frac{1-e^{2(L+1) x}}{1-e^{-2 x}}\right) \right] dr^2\simeq\frac{1}{8\pi} \int_{\sqrt{6 (K_c/K-1)}}^{\infty}\left[ \ln \left(\frac{1-e^{2(L+1) x}}{1-e^{-2 x}}\right) \right] dx^2 \nonumber\\
	&=& -\frac{1}{16\pi} \frac{2 \sqrt{x_t}\;  \text{Li}_2\left(-e^{-2\sqrt{x_t}}\right)+\text{Li}_3\left(-e^{-2\sqrt{x_t}}\right)}{(L+1)^2}+ \cdots,
\end{eqnarray}
where $R$ can be defined from the constraint  $(2\pi)\times (2\pi)=4\pi^2= \pi R^2$, i.e., $R=2\sqrt{\pi}$., and $(\cdots)$ stands for terms which are $L$-independent.

Summarizing the above, we conclude that the excess free energy can be written in a scaling form 
\begin{equation}
	\label{eq:excess-free-energy-scaling-DD}
	\beta \Delta f_{{\rm ex},3}^{({\rm DD})}(\beta,h=0)=\frac{1}{L^2} X_{\rm ex}^{(\rm DD)}(a_\beta t L^{1/\nu})
\end{equation}
where $a_\beta$ is a non-universal constants, and $X_{\rm ex}$ is an universal scaling function, $t=(T-T_c)/T_c$, where $T$ has the meaning of the temperature of the system, and $T_c$ is its bulk temperature. From \eq{eq:g1-int-r}, taking into account that with  $\nu=1/2$ one has $(2L+1)^2\simeq 4 L^2\simeq L^{1/\nu}$, we identify that 
\begin{equation}
	\label{eq:scaling-function-excess-free-energy-A}
	X_{\rm ex,GC}^{(\rm DD)}(x_t,h=0)=-\frac{1}{16 \pi} \left[2\sqrt{x_t}\; \text{Li}_2\left(e^{-2\sqrt{x_t}}\right)+\text{Li}_3\left(e^{-2\sqrt{x_t}}\right)\right]. 
\end{equation}

\subsection{Evaluation of the field dependent terms}
For 	Dirichlet-Dirichlet (DD) boundary conditions from \eq{eq:h-vector} and \eq{eq:DD} one has:
\begin{equation}
	\hat{h}_{\Lambda}^{({\rm DD})}({\mathbf k}) = \sum_{{\mathbf r}\in \Lambda}
	h({\mathbf r})\bar{u}_{\Lambda}^{({\rm DD})}({\mathbf r},{\mathbf k})=\sqrt{\frac{2 L_1 L_2}{L_3+1}}\; \delta_{k_1,0} \delta_{k_2,0}\; h\,\sum_{r=1}^{L_3} \sin \left[ r\;\varphi_{L_3}^{({\rm DD})}(k_3)\right], \varphi_{L_3}^{({\rm DD})}(k_3) = \pi \frac{k_3}{L_3+1}. 
	\label{eq:h-vector-DD}
\end{equation}
Setting $k_3=k, r_3=r$ and $L_3=L$, for a film geometry from \eq{eq:the-field-term} we arrive at 
\begin{equation}
	\label{eq:field-term-DD}
	P_{L}^{({\rm DD})}(\beta,h) = {2 h^2\over \beta L(L+1)}
	\sum_{k=1}^{L} {\left[\sum_{r=1}^{L} \sin \left[r\;\pi\frac{k}{L+1}\right]\right]^{2} \over  6 (K_c/K-1) + 2\left(1 -
		\cos \pi\frac{k}{L+1}\right)}.
\end{equation}

It is easy to show that
\begin{equation}
	\label{eq:the-sum-in-numerator-DD}
	\sum_{r=1}^{L} \sin \left(\pi\; r\frac{  k }{L+1}\right) = \left\{
	\begin{array}{cc}  \cot\left[\frac{\pi \; k}{2(L+1)}\right]& \quad \mbox{if\;} k \quad \mbox{is odd} \\
		0 & \quad \mbox{if\;} k \quad \mbox{is even} 
	\end{array}\right. .
\end{equation}
Thus, one has 
\begin{equation}
	\label{eq:the-field-term-summed-over-k}
	P_{L}^{({\rm DD})}(\beta,h) = { h^2\over \beta L(L+1)}
	\sum_{k=1}^{L} {\left[1-(-1)^k\right]\cot^2 \left(\frac{\pi }{2}\frac{ k}{L+1}\right)  \over  6 (K_c/K-1) + 2\left(1 -
		\cos \pi\frac{k}{L+1}\right) },
\end{equation}
and, therefore
\begin{equation}
	\label{eq:the-field-term-summed-over-k-M-DD}
	P_{L}^{({\rm DD})}(\beta,i \omega) = -{ \omega^2\over \beta L(L+1)}
	\sum_{k=1}^{L} {\left[1-(-1)^k\right]\cot^2 \left(\frac{\pi }{2}\frac{ k}{L+1}\right)  \over  6 (K_c/K-1) + 2\left(1 -
		\cos \pi\frac{k}{L+1}\right) }.
\end{equation}

Let us consider the small $k$ behavior of the above sum. One derives
\begin{eqnarray}
	\label{eq:the-field-term-small-k-DD-A}
	P_{L}^{({\rm DD})}(\beta,h) & \simeq & \frac{h^2}{\beta}{1\over L(L+1)}
	\sum_{k=1}^{L} { \left[1-(-1)^k\right]\over \left[\frac{\pi k}{2(L+1) }\right]^2 \left\{6 (K_c/K-1) +  \left[\frac{\pi k}{L+1 }\right]^2    \right\} } \nonumber\\
	&\simeq& \frac{8}{\pi^2} \frac{h^2}{\beta} 	{ (L+1)^3\over   L}
	\sum_{k=1}^{\infty} { 1\over (2k+1)^2 \left[x_t +\pi^2 (2k+1)^2\right]}\nonumber \\
	&=& \frac{ h^2}{\beta} 	{ (L+1)^3\over   L} \left\{\frac{1}{  x_t}\left[1-\frac{\tanh \left(\sqrt{x_t}/2\right)}{\sqrt{x_t}/2}\right]+{\cal O}(L^{-3})\right\}. 
\end{eqnarray}
In terms of the scaling variables, \eq{eq:the-field-term-small-k-DD-A} states that 
\begin{equation}
	\label{eq:field-term-scaling-form-DD}
	P_{L}^{({\rm DD})}(x_h,x_t)= L^{-3} X_P^{(\rm DD)}(x_t,x_h) \;\mbox{where} \; X^{(\rm DD)}_P(x_t,x_h)=\frac{x_h^2}{ x_t}\left[1-\frac{\tanh \left(\sqrt{x_t}/2\right)}{\sqrt{x_t}/2}\right]. 
\end{equation}

In the limits $x_t\to 0$ and $x_t\to\infty$ for the behavior of the field term one obtains
\begin{equation}
	\label{eq:as-P-small-and-large-y-DD}
	P_{L}^{({\rm DD})}(\beta,h) \simeq  \frac{h^2}{\beta} 	{ (L+1)^3\over   L} 
	\left \{\begin{array}{cc}
		1/12+{\cal O}(x_t), & x_t\to 0;\\
		& \\
		1/x_t+{\cal O}\left[\exp(-\sqrt{x_t})\right], &x_t\gg 1. 
	\end{array}\right.
\end{equation}
When $L\to \infty$, then $x_t \to \infty$, we obtain that 
\begin{equation}
	\label{eq:bulk-limit}
	\lim_{L\to \infty}	P_{L}^{({\rm DD})}(\beta,h)= \frac{h^2}{ 6\beta (K_c/K-1) }, 
\end{equation}
which indeed equals the bulk expression $P_\infty(\beta, h)$.

From \eq{eq:the-field-term-small-k-DD-A} for the behavior of the susceptibility in the finite system we derive
\begin{equation}
	\label{eq:susceptibility-DD-A}
	\chi_{L}^{({\rm DD})}(\beta,h)=\frac{2}{\beta} 	{ (L+1)^3\over   L} 
	\frac{1}{ x_t}\left[1-\frac{\tanh \left(\sqrt{x_t}/2\right)}{\sqrt{x_t}/2}\right].
\end{equation}

\subsection{Evaluation of the term dependent on magnetization}

For Dirichlet-Dirichlet (DD) boundary conditions along $z$ direction, i.e., $\tau =(p,p,{\rm DD})$,
having in mind \eq{eq:DD}, we obtain
	\begin{equation}
		\sum_{{\mathbf r}\in \Lambda}
		\bar{u}_{\Lambda}^{({\rm DD})}({\mathbf r},{\mathbf k})=2\sqrt{\frac{L_1 L_2}{L_3+1}}\; \delta_{k_1,0} \delta_{k_2,0}\;\sum_{r=1}^{L_3} \sin \left[r\;\varphi_{L_3}^{({\rm DD})}(k_3)\right],  \quad \mbox{with} \quad 
		\varphi_{L_3}^{({\rm DD})}(k_3) =  \frac{\pi k}{L_3+1}. 
		\label{eq:h-vector-ND-new}
	\end{equation}
It is easy to show that 
\begin{equation}
	\label{eq:sum-DD}
	2\sum_{r=1}^{L_3} \sin \left[r\;\varphi_{L_3}^{({\rm DD})}(k_3)\right]=\left[1-\left(-1\right)^{k_3}\right]\cot\left[\frac{\pi k_3}{2(1+L_3)}\right].
\end{equation}
With this, in the case of a film geometry we derive 
	\begin{eqnarray}
		\label{eq:the-M-term-DD}
		&& \frac{ K M^2/|\Lambda|}{
			\sum_{{\mathbf k}\in \Lambda} {[\sum_{{\mathbf r}\in \Lambda}
				\bar{u}_{\Lambda}^{\rm (DD)}({\mathbf r},{\mathbf k})]^{2} \over 2s/K - \mu_{\Lambda}^{\rm (DD)}({\mathbf k})}}= |\Lambda|\frac{ K m^2}{
			\sum_{{\mathbf k}\in \Lambda} {[\sum_{{\mathbf r}\in \Lambda}
				\bar{u}_{\Lambda}^{\rm (DD)}({\mathbf r},{\mathbf k})]^{2} \over 2s/K - \mu_{\Lambda}^{\rm (DD)}({\mathbf k})}}=L_3(L_3+1) \frac{ K m^2}{
			\sum_{k_3=1}^{L_3} {2\left[1-(-1)^{k_3}\right]\cot^2 \left(\frac{\pi }{2}\frac{  k_3}{L_3+1}\right) \over 2s/K - \mu_{\Lambda}^{\rm (DD)}({0,0,k_3})}}\\
		&=&L_3(L_3+1) \frac{ K m^2}{
			\sum_{k_3=1}^{L_3} {2\left[1-(-1)^{k_3}\right]\cot^2 \left(\frac{\pi }{2}\frac{  k_3}{ L_3+1}\right) \over 6(K_c/K-1) + 2\left(1 -
				\cos \pi\frac{k_3}{L_3+1}\right)} }\simeq  L_3(L_3+1)^{-3} \frac{\pi^2}{4} \frac{ K m^2}{
			\sum_{k=1}^{L_3} { 2\left[1-(-1)^{k_3}\right] \over k^2 \left[x_t +\pi^2 k^2\right]}}\nonumber \\
		&&= L_3(L_3+1)^{-3} { K m^2} \left\{\frac{1}{2 x_t}\left[1-\frac{\tanh \left(\sqrt{x_t}/2\right)}{\sqrt{x_t}/2}\right]\right\}^{-1} \nonumber\\ 
		&&={(L_3+1)}^{-3}2 x_t x_m^2\left[1-\frac{\tanh \left(\sqrt{x_t}/2\right)}{\sqrt{x_t}/2}\right]^{-1}, \nonumber
	\end{eqnarray}
where, now 
\begin{equation}
	\label{eq:scaling-regime}
	x_t= 6 (\beta_c/\beta-1) (L+1)^2 ={\cal O}(1) \quad \mbox{and} \quad x_m= m \sqrt{K L}.
\end{equation}
The last equation in 	\eq{eq:the-M-term-DD} implies that 
\begin{equation}
	\label{eq:Q-ND-final-A}
	Q_{L}^{\rm (DD)}(K,m;s)
	= {(L_3+1)}^{-3} 2 x_t x_m^2 \left[1-\frac{\tanh \left(\sqrt{x_t}/2\right)}{\sqrt{x_t}/2}\right]^{-1}.
\end{equation}
As we see, this term also perfectly fits in a scaling form. Subtracting now the bulk term given by 	\eq{eq:bulk-term-m-Helmhotz}, we obtain that the excess part of the free energy in $(T-m)$ ensemble with Dirichlet--Dirichlet boundary conditions is 
\begin{equation}
	\label{eq:excess-m-DD}
		Q_{\rm ex}^{\rm (DD)}(K,m;s|L)= L^{-2} 2 x_t x_m^2\left\{\left[1-\frac{\tanh \left(\sqrt{x_t}/2\right)}{\sqrt{x_t}/2}\right]^{-1}-1\right\}. 
\end{equation}
Therefore, the total excess free energy in the canonical ensemble is 
\begin{equation}
	\label{eq:scaling-function-excess-free-energy-A}
	X_{\rm ex,C}^{(\rm DD)}(x_t,x_m)=-\frac{1}{16 \pi} \left[2\sqrt{x_t}\; \text{Li}_2\left(e^{-2\sqrt{x_t}}\right)+\text{Li}_3\left(e^{-2\sqrt{x_t}}\right)\right]+2 x_t x_m^2 \left[1-\frac{\tanh \left(\sqrt{x_t}/2\right)}{\sqrt{x_t}/2}\right]^{-1}. 
\end{equation}
Thus, the contribution of $Q_{L}^{\rm (DD)}(K,m;s)$ term into the Helmholtz force is 
\begin{eqnarray}
	\label{eq:m-contribution-force-DD-A}
	\Delta \beta F_{\rm Helmholtz}^{\rm (DD)}(L,T,m) &=& -\frac{\partial}{\partial L} 	Q_{\rm ex}^{\rm (DD)}(K,m;s|L) \\
	&=& -L^{-3}\frac{4 x_m^2 x_t \left(x_t-2 \cosh \left(\sqrt{x_t}\right)+2\right)}{x_t-4 \sqrt{x_t} \sinh \left(\sqrt{x_t}\right)+(x_t+4) \cosh \left(\sqrt{x_t}\right)-4}. \nonumber 
\end{eqnarray}
With respect to the Helmholtz force, from \eq{eq:scaling-function-excess-free-energy-A} and 	\eq{eq:m-contribution-force-DD-A} we arrive at
\begin{eqnarray}
	\label{eq:helmholtz-DD-A}
	X_{\rm Helmholtz}^{(\rm DD)}(x_t,x_m) &=& -\frac{1}{8 \pi} \left[2\sqrt{x_t}\; \text{Li}_2\left(e^{-2\sqrt{x_t}}\right)+\text{Li}_3\left(e^{-2\sqrt{x_t}}\right)-2 x_t \ln\left(1-e^{-2\sqrt{x_t}}\right)\right] \\
	&& -\frac{4 x_m^2 x_t \left(x_t-2 \cosh \left(\sqrt{x_t}\right)+2\right)}{x_t-4 \sqrt{x_t} \sinh \left(\sqrt{x_t}\right)+(x_t+4) \cosh \left(\sqrt{x_t}\right)-4}. \nonumber
\end{eqnarray} 

\section{\textit{\textbf{The behavior of the interaction term, the field term, the excess free energy and the Casimir force under Neumann -- Dirichlet boundary conditions}}}
\label{sec:ND}

\subsection{Evaluation of the term dependent on magnetization}

For Neumann-Dirichlet boundary conditions along $z$ direction, i.e., $\tau =(p,p,{\rm ND})$ we have to calculate
$[\sum_{{\mathbf r}\in \Lambda}
\bar{u}_{\Lambda}^{\rm (ND)}({\mathbf r},{\mathbf k})]^{2}$, where we introduced the short-hand notation $\rm (ND)\equiv(p,p,{\rm ND}) $. 

Having in mind \eq{eq:DN}, we obtain
\begin{equation}
	\sum_{{\mathbf r}\in \Lambda}
	\bar{u}_{\Lambda}^{({\rm ND})}({\mathbf r},{\mathbf k})=2\sqrt{\frac{L_1 L_2}{2L_3+1}}\; \delta_{k_1,0} \delta_{k_2,0}\;\sum_{r=1}^{L_3} \cos \left[(r-1/2)\;\varphi_{L_3}^{({\rm ND})}(k_3)\right],  \quad \mbox{with} \quad 
	\varphi_{L_3}^{({\rm ND})}(k_3) = \pi \frac{2k_3-1}{2L_3+1},
	\label{eq:h-vector-ND-new}
\end{equation}
and 
\begin{equation}
	\label{eq:the-sum-in-numerator-new}
	2\sum_{r=1}^{L} \cos \left(\pi ( r-1/2)\frac{  (2 k-1) }{(2 L+1)}\right) =\frac{\sin \left(\pi \frac{ 2 k-1 }{2 L+1} L \right)}{\sin \left(\frac{\pi}{2} \frac{  2 k-1}{2 L+1}\right)}\quad \mbox{where}\quad \left[2\sum_{r=1}^{L} \cos \left(\pi ( r-1/2)\frac{  (2 k-1) }{(2 L+1)}\right)\right]^2 =\cot^2 \left(\frac{\pi }{2}\frac{ 2 k-1}{2 L+1}\right).
\end{equation}
Therefore, for a film geometry we obtain 
\begin{eqnarray}
	\label{eq:the-M-term}
	&& \frac{ K M^2/|\Lambda|}{
		\sum_{{\mathbf k}\in \Lambda} {[\sum_{{\mathbf r}\in \Lambda}
			\bar{u}_{\Lambda}^{\rm (ND)}({\mathbf r},{\mathbf k})]^{2} \over 2s/K - \mu_{\Lambda}^{\rm (ND)}({\mathbf k})}}= |\Lambda|\frac{ K m^2}{
		\sum_{{\mathbf k}\in \Lambda} {[\sum_{{\mathbf r}\in \Lambda}
			\bar{u}_{\Lambda}^{\rm (ND)}({\mathbf r},{\mathbf k})]^{2} \over 2s/K - \mu_{\Lambda}^{\rm (ND)}({\mathbf k})}}=L_3(2L_3+1) \frac{ K m^2}{
		\sum_{k_3=1}^{L_3} {\cot^2 \left(\frac{\pi }{2}\frac{ 2 k_3-1}{2 L_3+1}\right) \over 2s/K - \mu_{\Lambda}^{\rm (ND)}({0,0,k_3})}}\\
	&=&L_3(2L_3+1) \frac{ K m^2}{
		\sum_{k_3=1}^{L_3} {\cot^2 \left(\frac{\pi }{2}\frac{ 2 k_3-1}{2 L_3+1}\right) \over 6(K_c/K-1) + 2\left(1 -
			\cos \pi\frac{2k_3-1}{2L_3+1}\right)} } = L_3(2L_3+1)^{-3} \frac{\pi^2}{4} \frac{ K m^2}{
		\sum_{k=1}^{L_3} { 1\over (2k-1)^2 \left[x_t +\pi^2(2k-1)^2\right]}}\nonumber \\
	&&= L_3(2L_3+1)^{-3} { K m^2} \left\{\frac{1}{2 x_t}\left[1-\frac{\tanh \left(\sqrt{x_t}/2\right)}{\sqrt{x_t}/2}\right]\right\}^{-1}={(2L_3+1)}^{-3} 2 x_t x_m^2 \left[1-\frac{\tanh \left(\sqrt{x_t}/2\right)}{\sqrt{x_t}/2}\right]^{-1}. \nonumber
\end{eqnarray}
where  $x_m =\sqrt{K} m L^{\beta/\nu}= \sqrt{K} m L^{1/2}$, with  \cite[p. 359]{G92} $ \beta=1/4$ and $\nu=1/2$. The last equation implies that 
\begin{equation}
	\label{eq:Q-ND-final-A}
	Q_{L}^{\rm (ND)}(K,m;s)
	= {(2L_3+1)}^{-3} 2 x_t x_m^2 \left[1-\frac{\tanh \left(\sqrt{x_t}/2\right)}{\sqrt{x_t}/2}\right]^{-1}.
\end{equation}
As we see, this term perfectly fits in a scaling form. 

\section{\textit{\textbf{The behavior of the interaction term, the field term, the excess free energy and the Casimir force under Neumann -- Neumann boundary conditions}}}
\label{sec:NN}

\subsection{Evaluation of the interaction term}

For Neumann--Neumann boundary conditions,	from \eq{eq:U-term-NN} one explicitly obtains
	\begin{equation}
		\label{eq:uterm-explicit-DD}
		{\mathcal U}_{\rm L,GC}^{({\rm NN)}}(\beta)=	 {1\over L}
		\sum_{k=1}^{L}{\mathcal V}_{2}\left[ 6 (K_c/K-1) + 2\left(1-
		\cos \pi\frac{k-1}{L}\right)\right]=	\frac{1}{(2\pi)^2}\, \int_{-\pi}^{\pi} {\mathrm d}\theta_1\int_{-\pi}^{\pi} {\mathrm d}\theta_2\;	S^{({\rm NN})}(\beta,L|\theta_1,\theta_2),
	\end{equation}
with 
	\begin{equation}
		\label{eq:S-DD-expl}
		S^{({\rm NN})}(\beta,L|\theta_1,\theta_2)=\frac{1}{L}			\sum_{k=1}^{L}\,{\ln}\left[ 6 (K_c/K-1) + 2\sum_{\nu =1}^2
		(1-\cos\theta_{\nu}) +2\left(1 -
		\cos \pi\frac{k-1}{L}\right)\right].	
	\end{equation}

This sum is of the form
\begin{equation}
	\label{eq:definition-of-the-sum-NN}
	S^{({\rm NN})}(x,L)=\frac{1}{L}		\ln	\prod_{k=1}^{L}\,2\left[ \cosh(x) -
	\cos \pi\frac{k-1}{L}\right].
\end{equation}
The summations in \eq{eq:definition-of-the-sum-NN} can be performed using  the identity, which can be easily derived from \cite[1.431.3]{GR}:
\begin{equation}
	\label{eq:identity-DD}
	\prod_{k=1}^{L-1}\,2\left[ \cosh(x) -
	\cos \pi\frac{k}{L}\right]=\frac{\sinh\left[L x\right]}{\sinh(x)}. 
\end{equation}
With the help of the above identity one derives
\begin{equation}
	\label{eq:the-sum-DD}
	S^{({\rm NN})}(x,L)= \frac{1}{L} \ln \left[2 \tanh(x/2)\sinh\left(
	L x\right)\right]. 
\end{equation}
Obviously $	\lim_{L\to \infty} S^{({\rm NN})}(x,L)=x$.  Thus, the part of the excess free energy under Neumann -- Neumann
boundary conditions that depends only on the interaction term is 
\begin{eqnarray}
	\label{eq:excess-DD-slab-interaction}
	\beta \Delta f_{{\rm ex, GC}}^{({\rm NN})}(\beta,h=0)&=& \frac{1}{2}\, L\, \left[{\mathcal U}_{\rm L, GC}^{({\rm NN})}(\beta)- 	{\mathcal U}_{\infty}(\beta) \right] \\
	&=&  \frac{L}{8\pi^2}\, \int_{-\pi}^{\pi} {\mathrm d}\theta_1\int_{-\pi}^{\pi} {\mathrm d}\theta_2\;\left[ \frac{1}{L} \ln \left[2 \tanh(x/2)\sinh\left(
	L x\right)\right]-x\right] \nonumber\\
	&=& \frac{1}{8\pi^2}\, \int_{-\pi}^{\pi} {\mathrm d}\theta_1\int_{-\pi}^{\pi} {\mathrm d}\theta_2\;\left[ \ln \left(1-e^{-2 L x}\right) +\ln \left(\frac{1-\exp (-x)}{1+\exp (x)}\right)\right].\nonumber
\end{eqnarray}

Let us consider the behavior of $\beta \Delta f_{{\rm ex},3}^{({\rm NN})}(\beta,h=0)$ in the scaling regime 
\begin{equation}
	\label{eq:scaling-regime-DD}
	x_t= 6 (K_c/K-1) L^2 ={\cal O}(1).
\end{equation}
Obviously, if $x={\cal O}(1)$ then the $L$-dependent part of $\beta \Delta f_{{\rm ex},3}^{({\rm DD})}(\beta,h=0)$ will be exponentially small which respect to $L$ and will also contain ${\cal O}(1)$ $L$-independent terms. Thus, we need to consider the regime $2 L x={\cal O}(1)$. It follows that $x\ll 1$.   In the envisaged regime, using the same procedure as in \eq{eq:def-x} and \eq{eq:xl-r}, we arrive at 
	\begin{eqnarray}
		\label{eq:g1-int-r}
		\beta \Delta f_{{\rm ex, GC}}^{({\rm NN})}(\beta,h=0) & \simeq & \frac{1}{8\pi} \int_{0}^{R}\left[ \ln \left(1-e^{-2 L x}\right) +\ln \left(\frac{1-\exp (-x)}{1+\exp (-x)}\right)\right] dr^2\\
		& \simeq & \frac{1}{8\pi} \int_{\sqrt{6 (K_c/K-1)}}^{\infty}\left[ \ln \left(1-e^{-2 L x}\right) +\ln \left(\frac{1-\exp (-x)}{1+\exp (-x)}\right)\right] dx^2 \nonumber\\
		&=& -\frac{1}{16\pi} \frac{2 \sqrt{x_t}\;  \text{Li}_2\left(e^{-2\sqrt{x_t}}\right)+\text{Li}_3\left(e^{-2\sqrt{x_t}}\right)}{L^2}+ \cdots,
	\end{eqnarray}
where $R$ can be defined from the constraint  $(2\pi)\times (2\pi)=4\pi^2= \pi R^2$, i.e., $R=2\sqrt{\pi}$., and $(\cdots)$ stands for terms which are $L$-independent.

Summarizing the above, we conclude that the excess free energy can be written in a scaling form 
\begin{equation}
	\label{eq:excess-free-energy-scaling-DD}
	\beta \Delta f_{{\rm ex, GC}}^{({\rm NN})}(\beta,h=0)=\frac{1}{L^2} X_{\rm ex, GC}^{(\rm NN)}(a_\beta t L^{1/\nu})
\end{equation}
where $a_\beta$ is a non-universal constants, and $X_{\rm ex}$ is an universal scaling function, $t=(T-T_c)/T_c$, where $T$ has the meaning of the temperature of the system, and $T_c$ is its bulk temperature. From \eq{eq:g1-int-r}, taking into account that with  $\nu=1/2$ one has $L^2 \simeq L^{1/\nu}$, we identify that 
\begin{equation}
	\label{eq:scaling-function-excess-free-energy-A}
	X_{\rm ex,GC}^{(\rm NN)}(x_t)=-\frac{1}{16 \pi} \left[2\sqrt{x_t}\; \text{Li}_2\left(-e^{-2\sqrt{x_t}}\right)+\text{Li}_3\left(-e^{-2\sqrt{x_t}}\right)\right]. 
\end{equation}

\subsection{Evaluation of the field dependent terms}

For 	Neumann-Neumann (NN) boundary conditions, supposing again an homogeneous magnetic field, one has :
\begin{equation}
	u_{L}^{({\rm NN})}(r,k) =\left \{ \begin{array}{ll} L^{-1/2} & \mbox{for} \quad k=1 \\ \left[2/L\right]^{1/2}\cos (r-1/2)\varphi_{L}^{({\rm NN})}(k) & \mbox{for} \quad  k=2,\cdots, L \end{array}\right..
	\label{eq:NN2}
\end{equation}
and, therefore
\begin{eqnarray}
	\hat{h}_{\Lambda}^{({\rm NN})}({\mathbf k}) &=& \sum_{{\mathbf r}\in \Lambda}
	h({\mathbf r})\bar{u}_{\Lambda}^{({\rm NN})}({\mathbf r},{\mathbf k})=h \left\{
	\begin{array}{ll} \sqrt{L_1 L_2 L_3}\; \delta_{k_1,0} \delta_{k_2,0}\, &\mbox{for} \quad k_3=1   \\
		\sqrt{\frac{2 L_1 L_2}{L_3}}\; \delta_{k_1,0} \delta_{k_2,0} \sum_{r=1}^{L_3} \cos (r-1/2)\varphi_{L}^{({\rm NN})}(k_3), & \mbox{for} \quad k_3=2, \cdots, L_3.
	\end{array} \right. 
	\label{eq:h-vector-DD}
\end{eqnarray}
where $\varphi_{L}^{\rm (NN)}(k_3) = \pi(k_3-1)/L$. 
Setting $k_3=k, r_3=r$ and $L_3=L$, for a film geometry, from \eq{eq:the-field-term} and having in mind that
\begin{equation}
	\label{eq:NN-sum}
	\sum _{r=1}^L \cos \left[\frac{1}{L}\pi  (k-1) \left(r-\frac{1}{2}\right)\right]=\frac{\sin (\pi  k)}{2 \sin \left(\frac{\pi  (k-1)}{2 L}\right)} =\delta_{k,1},
\end{equation} 
we arrive at 
\begin{equation}
	\label{eq:field-term-NN}
	P_{L}^{({\rm NN})}(\beta,h) = {h^2\over \beta}
	{1 \over  6 (K_c/K-1) }= P_{L}^{({\rm per})}(K,h)=P_{\infty}^{({\rm per})}(K,h).
\end{equation}
Having in mind the results for periodic boundary conditions, we immediately conclude that 
the excess free energy related to the field term, see \eq{eq:free-energy-density-finite-region}, reads to
\begin{equation}
	\label{eq:excess-field-term-NN}
	\beta \Delta f_{{\rm ex},3}^{({\rm NN})}(\beta,h)= -\frac{1}{2} L \left[P_{L}^{({\rm NN})}(h;\beta) - P_{\infty}(h;\beta) \right]=0,
\end{equation}
i.e., it does \textit{not} depend on $h$. Then, the same is obviously true for the excess free energy and the Casimir force. 

From \eq{eq:field-term-NN} for the susceptibility we immediately obtain 
\begin{equation}
	\label{eq:susceptibility-per-NN}
	\beta 	\chi_{L}^{({\rm NN})}(\beta,h) =\beta 	\chi_{L}^{({\rm NN})}(\beta,h) = L^2 \;\frac{2}{ x_t}=  L^2 \;X^{(\rm NN)}_\chi( x_t),
\end{equation}
with
\begin{equation}
	\label{eq:scaling-susceptibility-per-NN}
	X^{(\rm NN)}_\chi( x_t)= X^{(\rm per)}_\chi( x_t)=\frac{2}{ x_t}.
\end{equation}

The behavior of the scaling function $X\chi^{(\rm NN)}(x_t)$ is given in Fig. \ref{fig:scaling-function-susceptibility-periodic}. 

\subsection{Evaluation of the term dependent on magnetization}

Similar to the result for periodic boundary conditions, in  the case of a film geometry, we derive 
	\begin{eqnarray}
		\label{eq:the-M-term-DD}
		&& \frac{ K M^2/|\Lambda|}{
			\sum_{{\mathbf k}\in \Lambda} {[\sum_{{\mathbf r}\in \Lambda}
				\bar{u}_{\Lambda}^{\rm (NN)}({\mathbf r},{\mathbf k})]^{2} \over 2s/K - \mu_{\Lambda}^{\rm (NN)}({\mathbf k})}}=  K m^2 \left[2d (K_c/K-1)\right].  \nonumber
	\end{eqnarray}
where, with $x_t$ and $x_m$ determined by 
\begin{equation}
	\label{eq:scaling-regime}
	x_t= 6 (\beta_c/\beta-1) L^2 ={\cal O}(1), \quad x_m= m \sqrt{K L},
\end{equation}
we arrive at 
\begin{equation}
	\label{eq:Q-ND-final}
	Q_{L}^{\rm (NN)}(K,m;s)
	= {L}^{-3} x_t x_m^2.
\end{equation}
As we see, this term also perfectly fits in a scaling form. It is, however equal to the corresponding term in the infinite system. Thus, the excess free energy in the canonical ensemble and, therefore, the Helmholtz force does not depend on $h$ under Neumann -- Neumann boundary conditions.

\section{\textit{\textbf{The behavior of the interaction term, the field term, the excess free energy, the Casimir force, the Helmholtz  and susceptibility under periodic boundary conditions in a lattice Gaussian model}}}
\label{sec:periodic}

We start from \eq{eq:U-term-per}.
\begin{equation}
	{\mathcal U}_{L,3}^{(p)}(K)=   {1\over L}
	\sum_{k=1}^{L}{\mathcal V}_{2}\left[ 6 (K_c/K-1) + 2\left(1-
	\cos 2\pi\frac{k}{L}\right)\right]=\frac{1}{(2\pi)^2}\, \int_{-\pi}^{\pi} {\mathrm d}\theta_1\int_{-\pi}^{\pi} {\mathrm d}\theta_2\;	S^{(p)}(\beta,L|\theta_1,\theta_2),
	\label{eq:U-term-per}
\end{equation}		
where 
\begin{equation}
	\label{eq:S-p-expl}
	S^{(p)}(\beta,L|\theta_1,\theta_2)=\frac{1}{L}			\sum_{k=1}^{L}\,{\ln}\left[ 6 (K_c/K-1) + 2\sum_{\nu =1}^2
	(1-\cos\theta_{\nu}) +2\left(1 -
	\cos 2\pi\frac{k}{L}\right)\right].	
\end{equation}
The sum in \eq{eq:S-p-expl} is of the form
\begin{equation}
	\label{eq:definition-of-the-sum-pbc}
	S^{(p)}(x,L)=\frac{1}{L}		\ln	\prod_{k=0}^{L-1}\,2\left[ \cosh(x) -
	\cos 2\pi\frac{k}{L}\right],
\end{equation}
where $x=x(\beta|\theta_1,\theta_2)$ is defined as 
\begin{equation}
	\label{eq:def-x}
	\cosh x=1+3 (K_c/K-1) + \sum_{\nu =1}^2
	(1-\cos\theta_{\nu}). 
\end{equation}
This sum can be easily calculated using the identity \cite[1.395.2]{GR}
\begin{equation}
	\label{eq:identity-periodic}
	\prod_{k=0}^{L-1}\,2\left[ \cosh(x) -
	\cos 2\pi\frac{k}{L}\right]=2\left[\cosh(Lx)-1\right]
\end{equation}
With the help of the identity \eq{eq:identity-periodic} one derives 
\begin{equation}
	\label{eq:the-sum-periodic}
	S^{(p)}(x,L)= \frac{1}{L} \ln 2\left[\cosh(Lx)-1\right]. 
\end{equation}
Obviously $	\lim_{L\to \infty} S^{(p)}(x,L)=x$.  Thus, the part of the excess free energy under periodic boundary conditions that depends only on the interaction term is 
\begin{eqnarray}
	\label{eq:excess-ND-slab-interaction}
	\beta \Delta f_{{\rm ex},3}^{(p)}(\beta,h=0)&=& \frac{1}{2}\, L\, \left[{\mathcal U}_{L,3}^{(p)}(\beta)- 	{\mathcal U}_{\infty,3}(\beta) \right] \nonumber\\
	&=&  \frac{L}{8\pi^2}\, \int_{-\pi}^{\pi} {\mathrm d}\theta_1\int_{-\pi}^{\pi} {\mathrm d}\theta_2\;\left[ \frac{1}{L} \ln 2\left[\cosh(Lx)-1\right]-x\right] \\
	&=& \frac{1}{8\pi^2}\, \int_{-\pi}^{\pi} {\mathrm d}\theta_1\int_{-\pi}^{\pi} {\mathrm d}\theta_2\;\ln \left[1- 2  \exp{(-L x)} + \exp{(-2  L  x)}  \right]. \nonumber
\end{eqnarray}
Obviously, if $x={\cal O}(1)$ then $\beta \Delta f_{{\rm ex},3}^{(p)}(\beta,h=0)$ will be exponentially small. Thus, we need to consider the regime $L x={\cal O}(1)$. It follows that then $x \ll 1$.  From \eq{eq:def-x} we obtain
\begin{equation}
	\label{eq:x-expansion}
	1+\frac{1}{2} x^2=3 (K_c/K-1) +\frac{1}{2}\left(\theta_1^2+\theta_2^2\right).
\end{equation}
It follows that 
\begin{equation}
	\label{eq:xl-r}
	x^2=6 (K_c/K-1) +\left(\theta_1^2+\theta_2^2\right)=6 (K_c/K-1) +r^2,
\end{equation}
where we have introduced polar coordinates. 

Let us consider the behavior of $\beta \Delta f_{{\rm ex},3}^{(p)}(\beta,h=0)$  in the scaling regime 
\begin{equation}
	\label{eq:scaling-regime-A}
	x_t= 6 (K_c/K-1) L^2 ={\cal O}(1).
\end{equation}
Then $\beta \Delta f_{{\rm ex},3}^{(p)}(\beta,h=0)$ becomes
\begin{eqnarray}
	\label{eq:g1-int-r}
	\beta \Delta f_{{\rm ex},3}^{(p)}(\beta,h=0) & \simeq & \frac{1}{8\pi} \int_{0}^{R}\ln \left[1- 2  \exp{(-L x)} + \exp{(-2  L  x)}  \right] dr^2\\
	&\simeq&\frac{1}{8\pi} \int_{\sqrt{6 (K_c/K-1)}}^{\infty}\ln \left[1- 2  \exp{(-L x)} + \exp{(-2  L  x)}  \right] dx^2 \nonumber\\
	&=& -\frac{1}{2\pi} \frac{\sqrt{x_t}\; \text{Li}_2\left(e^{-\sqrt{x_t}}\right)+\text{Li}_3\left(e^{-\sqrt{x_t}}\right)}{L^2},
\end{eqnarray}
where $R$ can be defined from the constraint  $(2\pi)\times (2\pi)=4\pi^2= \pi R^2$, i.e., $R=2\sqrt{\pi}$.

Summarizing the above, we conclude that the excess free energy can be written in a scaling form 
\begin{equation}
	\label{eq:excess-free-energy-scaling-per}
	\beta \Delta f_{{\rm ex},3}^{(p)}(\beta,h=0)=\frac{1}{L^2} X_{\rm ex}^{(p)}(a_\beta t L^{1/\nu})
\end{equation}
where $a_\beta$ is a non-universal constants, and $X_{\rm ex}$ is an universal scaling function, $t=(T-T_c)/T_c$, where $T$ has the meaning of the temperature of the system, and $T_c$ is its bulk temperature. From \eq{eq:g1-int-r}, taking into account that with  $\nu=1/2$ one has $L^2 = L^{1/\nu}$, we identify that 
\begin{equation}
	\label{eq:scaling-function-excess-free-energy}
	X_{\rm ex}^{(p)}(x_t)=-\frac{1}{2 \pi} \left[\sqrt{x_t}\; \text{Li}_2\left(e^{-\sqrt{x_t}}\right)+\text{Li}_3\left(e^{-\sqrt{x_t}}\right)\right]<0. 
\end{equation}

This leads to
\begin{equation}
	\label{eq:Casimir-periodic-zero-field}
	X_{\rm Casimir}^{(p)}(x_t)=	-\frac{1}{\pi}\left[\text{Li}_3\left(e^{-\sqrt{\text{xt}}}\right)+\sqrt{\text{xt}} \text{Li}_2\left(e^{-\sqrt{\text{xt}}}\right)-\frac{1}{2} \text{xt} \log \left(1-e^{-\sqrt{\text{xt}}}\right)\right].
\end{equation}

and 
\begin{equation}
	\label{eq:Cas-amplitude-periodic}
	\Delta^{(p)}_{\rm Casimir}\equiv X_{\rm ex}^{(p)}(x_t=0)=	-\frac{\zeta (3)}{2\pi }. 
\end{equation}

The dependence of the free energy on the field variable is given by the "field term", given by \eq{eq:the-field-term}. 
For a homogeneous filed $h$ and for periodic boundary  conditions ${\rm (per)}\equiv (p,p,p)$,  from \eq{eq:per} and 	\eq{eq:angles} it is easy to obtain that 
\begin{equation}
	\hat{h}_{\Lambda}^{(\rm per)}({\mathbf k}) = \sum_{{\mathbf r}\in \Lambda}
	h({\mathbf r})\bar{u}_{\Lambda}^{({\rm (per)})}({\mathbf r},{\mathbf k})=\sqrt{L_1 L_2 L_3}\; \delta_{k_1,0} \delta_{k_2,0} \delta_{k_3,0}\; h,
	\label{eq:h-vector-per}
\end{equation}
and 
\begin{equation}
	\label{eq:P-per}
	P_{L}^{({\rm per})}(K,h) 	=\frac{h^2}{6 \beta (K_c/K-1)} =P_{\infty}(K,h) = \lim_{L\to \infty} P_{L}^{({\rm per})}(K,h). 
\end{equation}	
Then for the excess free energy related to the field term, see \eq{eq:free-energy-density-finite-region}, one derives
\begin{equation}
	\label{eq:excess-field-term-per}
	\beta \Delta f_{{\rm ex},3}^{({\rm per})}(\beta,h)= -\frac{1}{2} L \left[P_{L}^{({\rm per})}(h;\beta) - P_{\infty}(h;\beta) \right]=0.
\end{equation}
\eq{eq:P-per} can be written in a scaling form 
\begin{equation}
	\label{eq:P-per-scaling}
	P_{L}^{({\rm per})}(K,h) 	=\frac{h^2}{6 \beta (K_c/K-1)} = \frac{1}{L^3}\frac{x_h^2}{x_t}\equiv  X_P^{(\rm per)}(x_t,x_h), \quad \mbox{with} \quad X_P^{(\rm per)}(x_t,x_h)=\frac{x_h^2}{x_t}.
\end{equation}
where $x_h=\beta^{-1/2}L^{5/2}\; h$ and $x_t= 6 (K_c/K-1) L^2$. 

\section{Partition function  and Casimir force of the continuum Gaussian model with Neumann-Dirichlet boundary conditions} \label{app:Neumann-Dirichlet}

\begin{figure}[htbp]
\begin{center}
\includegraphics[width=5in]{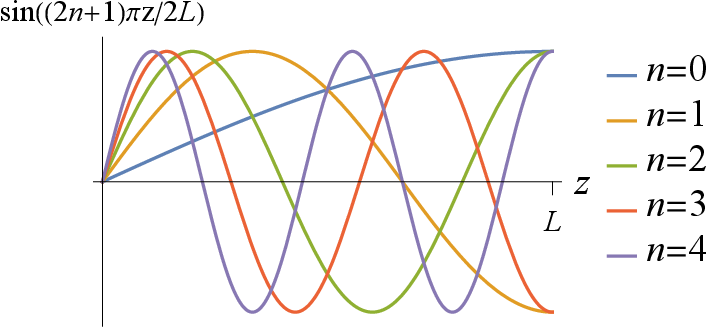}
\caption{The functions in (\ref{eq:3dg4p1})}
\label{fig:DNfunctions}
\end{center}
\end{figure}

The calculations in the case of Dirichlet-Dirichlet boundary conditions point the way to evaluating the partition function and the Casimir force of the case of Neumann-Dirichlet boundary conditions. 
In this case the (unnormalized) basis functions are, exclusive of their dependence on the in-plane coordinates,
\begin{equation}
\sin\left[\frac{\pi(2n+1)}{2L}  z\right] \label{eq:3dg4p1}
\end{equation}
Examples of these functions are shown in Fig. \ref{fig:DNfunctions}
Focusing on the $h$-independent contribution to the partition function, the sum to perform in this case is (see (\ref{eq:prelim7}))
\begin{equation}
\frac{1}{2}\sum_{n=0}^{\infty} \frac{1}{t+Q^2 +((2n+1) \pi/2L)^2} = \frac{L \tanh \left(L \sqrt{t+Q^2}\right)}{4 \sqrt{t+Q^2}}  \label{eq:3dg4p2}
\end{equation}
Note that in the limit of large $L$ the right hand side goes to the expected asymptotic form.
If we subtract that limiting form, and integrate with respect to $t$,  we are left with
\begin{equation}
\frac{1}{2}\left(\log \left(\cosh \left(L \sqrt{t+Q^2}\right)\right)-L \sqrt{t+Q^2}\right) \label{eq:3dg4p3}
\end{equation}
Finally, we take minus the derivative of this with respect to $L$, leaving us with
\begin{equation}
\frac{1}{2}\left(-\sqrt{r+Q^2} \tanh \left(L \sqrt{t+Q^2}\right)+\sqrt{t+Q^2} \right)= \sqrt{t+Q^2} \frac{e^{-L\sqrt{t+Q^2}}}{e^{L\sqrt{t+Q^2}}+e^{-L\sqrt{t+Q^2}}}  \label{eq:3dg4p4}
\end{equation}
Making use of the analysis of previous sections, this leaves us with the following result for the Casimir force in the case of the $d$-dimensional Gaussian model with Neumann-Dirichlet boundary conditions
\begin{eqnarray}
\lefteqn{\frac{\mathcal{K}_{d-1}}{(2 \pi)^{d-1}} \int_0^{\infty}Q^{d-2}\sqrt{t+Q^2} \frac{e^{-L\sqrt{t+Q^2}}}{e^{L\sqrt{t+Q^2}} + e^{-L\sqrt{t+Q^2}}} dQ} \nonumber \\ 
&= & \frac{\mathcal{K}_{d-1}}{(2 \pi)^{d-1}} \frac{1}{L^d} ( x_t)^{d/2}\int_0^{\infty} w^{d-2} \sqrt{1+w^2} \frac{e^{-\sqrt{x_t}\sqrt{1+w^2}}}{e^{\sqrt{x_t}\sqrt{1+w^2}}+e^{-\sqrt{x_t}\sqrt{1+w^2}}} dw \nonumber \\ & = & \frac{1}{L^d} X_{\rm Cas, D/N}^{(d)}(x_t) \label{eq:3dg4p5}
\end{eqnarray}
When $d=3$, we have
\begin{equation}
X_{\rm {Cas},D/N,I}^{(3)}(x_t)= -\frac{2 \sqrt{x_t} \text{Li}_2\left(-e^{-2 \sqrt{x_t}}\right)+\text{Li}_3\left(-e^{-2
   \sqrt{x_t}}\right)-2 x_t \log \left(e^{-2 \sqrt{x_t}}+1\right)}{8 \pi } 
\end{equation}

%Figure \ref{fig:DNplot1} shows what the function $X_{\rm Cas, D/N,I}^{(d)}(x_t)$ looks like when $d=3$. 
%\begin{figure}[htbp]
%\begin{center}
%\includegraphics[width=5in]{DNplot1.eps}
%\caption{The function $X_{\rm Cas, D/N}^{(3)}(x_t)$, as given in (\ref{eq:3dg4p5}).}
%\label{fig:DNplot1}
%\end{center}
%\end{figure}

In order to find the $h$-dependent contribution to the Casimir force we turn to the normalized the basis set in the case of Neumann-Dirichlet boundary conditions. Assuming that the boundary conditions are Dirichlet at $z=0$ and Neumann at $z=L$, this basis set is 
\begin{equation}
\psi^{(n)}_{DN}(z) = \sqrt{\frac{2}{LA}} \sin \left((n+1/2) \pi z/L \right) \label{eq:dnh1}
\end{equation}
where $A$ is the in-plane area of the slab, with $n$ an integer and
\begin{equation}
0 \leq n <\infty \label{eq:dnh2}
\end{equation} 
It is straightforward to establish that
\begin{equation}
\int_0^L \psi_{DN}^{(n)}(z)^2 \, dz \, d^{d-1}x = 1 \label{eq:dnh3}
\end{equation}
where $x$ is the $d-1$-dimensional in plane coordinate system,
while
\begin{equation}
\int_0^L \psi_{DN}^{(n)}(z) \, dz =  \frac{2\sqrt{2LA}}{(2n+1) \pi} \label{eq:dnh4}
\end{equation}

As it turns out there is no need to take into account any dependence of the basis set on coordinates in the plane of the slab. This is because a constant ordering field couples only to order parameter configurations that are independent of those coordinates. With this in mind, we expand the order parameter as follows
\begin{equation}
\Psi(z) = \sum_{n=0}^{\infty} a_n^{(DN)} \psi_{DN}^{(n)}(z) \label{eq:dnh5}
\end{equation}
The Gaussian integrations over the $a_n^{(DN)}$'s leaves us with the summation over $n$ for the $h$-dependent contribution to the grand partition function
\begin{eqnarray}
\lefteqn{\exp \left[h^2 \sum_{n=0}^{\infty}\left(\frac{2 \sqrt{2LA}}{(2n+1) \pi} \right)^2\frac{1}{4(( \pi (n+1/2)/L)^2 +t)} \right]} \nonumber \\  &=& \exp \left[ A h^2 \left(\frac{L}{4t} - \frac{\tanh(L\sqrt{t})}{4t^{3/2}} \right)\right] \nonumber \\ & = &  \exp \left[ \frac{Ah^2}{4t^{3/2}}\left( L \sqrt{t} - \tanh(L \sqrt{t})\right)\right] \label{eq:dnh6}
\end{eqnarray}
where the evaluation of the sum over $n$ in (\ref{eq:dnh6}) is accomplished with the use of (\ref{eq:prelim7}) and a partial fraction decomposition of the summand. 
The first term in parentheses on the last line of (\ref{eq:dnh6}) gives us exactly the same expression as the $h$-dependent contribution to the partition function of the slab with periodic boundary conditions. Its influence on the Casimir force is exactly canceled by the influence of the bulk. What remains is 
\begin{eqnarray}
-h^2 \frac{\partial}{\partial L} \tanh(L \sqrt{t}) /4t^{3/2} &=&-\frac{h^2}{4t} {\mathop{\rm sech}}^2(L\sqrt{t}) \nonumber \\ & = & -\frac{h^2L^2}{4x_t}{\mathop{\rm sech}}^2(\sqrt{x_t}) \nonumber \\
& = & -\frac{1}{L^d} \frac{x_h^2}{4x_t} {\mathop{\rm sech}}^2(\sqrt{x_t})  \label{eq:dnh7}
\end{eqnarray}
where we have made use of the definition of the scaling combination $x_h$ in (\ref{eq:prelimb}).

The total scaling function $X_{\rm{Casimir}}^{(ND)}(x_t,x_h)$ is given by
\begin{eqnarray}
\lefteqn{X_{\rm{Casimir}}^{(ND)}(x_t,x_h)} \nonumber \\ &=-&\frac{2 \sqrt{x_t} \text{Li}_2\left(-e^{-2 \sqrt{x_t}}\right)+\text{Li}_3\left(-e^{-2
   \sqrt{x_t}}\right)-2 x_t \log \left(e^{-2 \sqrt{x_t}}+1\right)}{8 \pi }\nonumber \\ && - \frac{x_h^2}{4x_t} {\mathop{\rm sech}}^2(\sqrt{x_t}) \label{eq:dnh8}
\end{eqnarray}

%For the behavior of the Casimir and Helmholtz forces under Neumann-Dirichlet boundary conditions, see Figs. \ref{fig:Casimir-ND} and \ref{fig:Helmholtz-ND}.

%Figure \ref{fig:totalX} shows what this function looks like.
%\begin{figure}[htbp]
%\begin{center}
%\includegraphics[width=5in]{totalX.eps}
%\caption{The total scaling contribution to the Casimir force for Neumann-Dirichlet boundary conditions in the three dimensional Gaussian model with a scalar order parameter,$X_{D/N }^{(3)}(x_t,x_h)$. Note that this function can be both positive (repulsive) and negative (attractive).}
%\label{fig:totalX}
%\end{center}
%\end{figure}

%\pagebreak

\subsection{The evaluation of the $M$-dependence of the Helmholtz force of the continuum Gaussian model with Neumann-Dirichlet boundary conditions}

We start with the $h$-dependent contribution to the grand partition function, as shown in (\ref{eq:dnh6}. We multiply this expression by $e^{- i h M}$ and integrate over $h$. Setting aside an overall factor of $1/2 \pi$, needed for the definition of the delta function, and pulling out the factor that explicitly depends on $M$, and then dividing by the cross-sectional area of the slab, $A$ we are left with the following expression for the the free energy per unit area of the system. 
\begin{equation}
\exp \left[ -\frac{M^2}{A} t^{3/2}/(L\sqrt{t}-\tanh(L \sqrt{t}) )\right] =\frac{m^2Lt}{1-\tanh(L\sqrt{t})}\label{eq:helm1}
\end{equation}
where we have used the relationship between the total magnetization, $M$ and the magnetization per unit volume, $m$
\begin{equation}
M=mAL \label{eq:helm2}
\end{equation}
The free energy per unit area of the bulk is
\begin{equation}
m^2(L_0-L) t \label{eq:helm2}
\end{equation}
where $L_0$ is the total extent of the bulk region. 
Adding the two we are left with 
\begin{equation}
\frac{m^2Lt}{1-\tanh(L\sqrt{t})/L\sqrt{t}} +m^2(L_0-L)
\end{equation}

Taking minus the $L$-derivative of the above expression, multiplying by $L^d$ and expressing $L$ and $t$ in terms of their scaling counterparts (see (\ref{eq:prelima}) and (\ref{eq:prelimc})) we have for the scaling contribution to the Helmholtz force of this system
\begin{equation}
X_{\rm Helmholtz}^{(DN)}=\frac{ x_m^2 x_t \left(\sinh ^2\left(\sqrt{x_t}\right)-x_t\right)}{\left(\sinh
   \left(\sqrt{x_t}\right)-\sqrt{x_t} \cosh \left(\sqrt{x_t}\right)\right){}^2}
\end{equation}

For the behavior of the Casimir and Helmholtz forces under Neumann-Dirichlet boundary conditions, see Figs. \ref{fig:Casimir-ND} and \ref{fig:Helmholtz-ND}.

\section{Partition function and Casimir force of the continuum Gaussian model with periodic boundary conditions}
\label{app:periodic}

In the case of periodic boundary conditions, we make use of the following orthonormal set of eigenfunctions. 

\begin{eqnarray}
\psi_c^{(n)}(z) &=&\sqrt{2/L}\cos(2 \pi n z/L)  \label{eq:periodic1} \\
\psi_s^{(n)}(z) & = & \sqrt{ 2/L }\sin(2 \pi n z/L)  \label{eq:periodic2} \\
\psi_0(z) & = & \sqrt{1/L} \label{eq:periodic3}
\end{eqnarray}
with $n$ a positive integer.
It is straightforward to  show that this set is orthonormal as a function of $z$ in that 
\begin{eqnarray}
\int_0^L \psi_c^{(n)}(z) \psi_c^{(m)}(z) \, dz &=& \delta_{m,n} \label{eq:periodic4} \\
\int_0^L \psi_s^{(n)}(z) \psi_s^{(m)}(z) \, dz & = & \delta_{m,n} \label{eq:periodic5} \\
\int_0^L \psi_0(z)^2 \, dz & = & 1  \label{eq:periodic6}
\end{eqnarray}
In the case of higher dimensions, we construct a new basis set by multiplying the functions (\ref{eq:periodic1})--(\ref{eq:periodic3}) by  suitable functions of the orthogonal position variables. Those functions can be taken to be of the form $e^{i \vec{Q} \cdot \vec{R}}$, where $\vec{R}$ is a $d-1$-dimensional position vector in the plane of the slab and $\vec{Q}$ is in its reciprocal space.

We then express the order parameter as follows
\begin{equation}
\psi(z,\vec{R}) = \sum_{\vec{Q}}e^{i \vec{Q}\cdot \vec{R}}\left(\sum_{n=1}^{\infty} a_n^{(c)} \psi_c^{(n)}(z) + \sum_{n=1}^{\infty} a_n^{(s)} \psi_c^{(n)}(z) + a_0 \psi_0(z) \right) \label{eq:periodic7}
\end{equation}

The free energy for a given configuration of the Gaussian order parameter, in terms of the amplitudes in the expansion of the order parameter in the basis set (\ref{eq:periodic4})--(\ref{eq:periodic6}), is 

\begin{equation}
\sum_{\vec{Q}} \left[\sum_{n=1}^{\infty}(a_n^{(c)\, 2}  +a_n^{(s) \, 2 } )\left(t+Q^2+(2 \pi n/L)^2 \right) + a_0^2 t  -ha_0 \sqrt{L} \right]  \label{eq:periodic8}
\end{equation}
The last term in brackets above reflects the fact that the only basis function that the constant external field couples to is the constant function in (\ref{eq:periodic3})

The next step is to exponentiate the expression in (\ref{eq:periodic8}), multiply by either $-1/\beta$, or setting $\beta=1$, by -1, and, after that, to perform the Gaussian integrals over the $a_n^{(c)}$'s, the $a_n^{(s)}$'s, and $a_0$. 
The resulting partition function is given by
\begin{eqnarray}
\lefteqn{\mathcal{Z}} \nonumber \\ &=&\exp \left[\frac{1}{\beta}\left( \frac{h^2 LA}{4t}+\frac{A}{(2 \pi)^{d-1}}\int \, d^{d-1}Q\sum_{n=-\infty}^{\infty} \frac{1}{2} \ln \left( \frac{t+Q^2 +(2 \pi n/L)^2}{\pi}\right)  \right) \right] \nonumber \\ \label{eq:periodic9}
\end{eqnarray}
The coefficient $A$ in (\ref{eq:periodic9}) is the $d-1$ dimensional area of the slab. 

As our next step we evaluate the sum over $n$ on the right hand side of the expression for the partition function. To achieve this, we take the $t$-derivative of the logarithm of the summand, perform the sum over $n$ and then integrate the resulting expression with respect to $t$. Taking the derivative of the summand in (\ref{eq:periodic9}) with respect to $t$ leaves us with the sum
\begin{eqnarray}
\frac{1}{2} \sum_{n=-\infty}^{\infty} \frac{1}{t+Q^2 +(2 \pi n/L)^2} = \frac{L \coth \left(\frac{1}{2} L \sqrt{Q^2+t}\right)}{4 \sqrt{Q^2+t}}   \label{eq:periodic10}
\end{eqnarray}
which follows from (\ref{eq:prelim6}). This integrates up to
\begin{eqnarray}
 \mathop{\rm ln} \left(\sinh \left( L \sqrt{Q^2+t}/2\right)\right) \label{eq:periodic11}
\end{eqnarray}
The large-$L$ limit of (\ref{eq:periodic11}) is 
\begin{equation}
L\sqrt{t+Q^2}/2 \label{eq:periodic12}
\end{equation}

To find the contribution to the Casimir force per unit area, we  take the $L$-derivative of the difference between (\ref{eq:periodic12}) and (\ref{eq:periodic11}) and then integrate over $\vec{Q}$. The derivative yields
\begin{equation}
\frac{\sqrt{Q^2+t}}{1-e^{L \sqrt{Q^2+t}}} \label{eq:periodic13}
\end{equation}
The sum over values of $\vec{Q}$ is expressible as an integral. Introducing the variable $w=QL$ and writing $t=x_t/L^2$ (recall the definitions at the beginning of this section) we find for the Casimir force at $h=0$
\begin{equation}
f_{\rm Casimir}^{(p)}(L,x_t) = \frac{1}{L^d} \int_0^{\infty}\frac{\mathcal{K}_{d-1}}{(2 \pi)^{d-1}} w^{d-2} \frac{\sqrt{w^2+x_t}}{1-\exp[\sqrt{w^2+x_t}]} dw \label{eq:genres1}
\end{equation}

 Setting $d=3$ and then performing the calculation, we end up with
\begin{eqnarray}
%\lefteqn{-\frac{K_{d-1}}{(2 \pi)^{d-1}} \int_0^{\infty} Q^{d-2} \sqrt{Q^2+t} \frac{e^{-L\sqrt{Q^2+t}}}{e^{L\sqrt{Q^2+t}}-e^{-L\sqrt{Q^2+t}}} \, dQ} \nonumber \\
%& = & \frac{1}{L^d} \left(-\frac{K_{d-1}}{(2 \pi)^{d-1}}x_t^{d/2} \int_0^{\infty} w^{d-2} \sqrt{1+w^2}\frac{e^{-\sqrt{x_t(1+w^2)}}}{e^{\sqrt{x_t(1+w^2)}}-e^{-\sqrt{x_t(1+w^2)}}} \, dw \right) \nonumber \\ 
%& = & \frac{1}{L^d}  X^{(\rm{per},3)}_{I}(x_t)   \label{eq:periodic14}
\lefteqn{\int_0^{\infty}\frac{\mathcal{K}_2}{(2 \pi)^2}Q\frac{\sqrt{Q^2+t}}{1-\exp(L \sqrt{Q^2+t})}dQ} \nonumber \\
& = & \frac{-2 \sqrt{x_t} \text{Li}_2\left(e^{-\sqrt{x_t}}\right)-2
   \text{Li}_3\left(e^{-\sqrt{x_t}}\right)+x_t \log \left(1-e^{-\sqrt{x_t}}\right)}{2 \pi
    L^3} \label{eq:final1}
\end{eqnarray}
where, to get to the last line of (\ref{eq:final1}) we  made use of the definition (\ref{eq:prelim4}) of $x_t$. The clear implication of  (\ref{eq:genres1}) and (\ref{eq:final1}) is that we can express the $h=0$ contribution to the Casimir force as $L^{-d}$---where $d=3$---times a function of the scaling temperature variable $x_t$. 
The coefficient $\mathcal{K}_2$ in the equations above is the  the $d=2$ instance of the geometric factor
\begin{equation}
\mathcal{K} _d = \frac{2 \pi^{d/2}}{\Gamma\left( \frac{d}{2}\right)}  \label{eq:periodic15}
\end{equation} 
From this we find 
\begin{equation}
X_{\rm{Casimir}}^{(p)}(x_t)=-\frac{2 \sqrt{x_t} \text{Li}_2\left(e^{-2 \sqrt{x_t}}\right)+\text{Li}_3\left(e^{-2
   \sqrt{x_t}}\right)-2 x_t \log \left(1-e^{-2 \sqrt{x_t}}\right)}{8 \pi } \label{eq:periodic16}
\end{equation}
where $\mathop{\rm{Li}}_j(x)$ is the polylogarithm function; see \cite{GR}. 

%A plot of the function $X_{\rm{Casimir}}^{(p)}(x_t)$ is shown in Fig. \ref{fig:periodicplot1}.
%\begin{figure}[htbp]
%\begin{center}
%\includegraphics[width=5in]{periodicplot2a.eps}
%\caption{The function $X_{\rm{Casimir}}^{(p)}(x_t)$, plotted versus $x_t$; See (\ref{eq:periodic16})}
%\label{fig:periodicplot1}
%\end{center}
%\end{figure}

The first term in parentheses in Eq. (\ref{eq:periodic9}) gives us the $h$-dependent contribution to the free energy: $-h^2LA/4t$. This is to be compared to the corresponding free energy of a neighboring bulk phase, which goes as $-h^2(L_0-L)A/4t$, where $L_0$ is an extent that will ultimately be taken to go to infinity. If you add the two free energies, the dependence on $L$, the thickness of the slab, disappears. This means that there is no $h$-dependent free energy when slab  boundary conditions are periodic, and hence no $h$-dependent contribution to the Casimir force. The lack of dependence on $h$ of the Casimir force leads directly to a lack of dependence on $M$ of the Helmholtz force. In fact,  for periodic boundary conditions the Casimir force and the Helmholtz forces are identical.
For the behavior of the Casimir and Helmholtz forces for periodic boundary conditions, see Fig. \ref{fig:sscaling-function-Casimir-periodic-zero-field}.

\section{Casimir and Helmholtz forces in the continuum Gaussian model with Neumann-Neumann boundary conditions} \label{app:Neumann-Neumann}

This case is straightforward given results obtained previously. The ``interaction'' contribution to the fluctuation-induced force is precisely the same as in the case of Dirichlet-Dirichlet boundary conditions, and the $h$ and $m$ dependent contributions to the Casimir and Helmholtz forces are equal to zero.  The reasoning is as follows:
In the case of Neumann-Neumann boundary conditions, the relevant eigenfunctions are $\phi_n(z)$, where $z$ ranges from 0 to $L$, and 
\begin{equation}
\phi_n(z) = \left\{ \begin{array}{ll} \sqrt{1/L} & n=0 \\ \sqrt{2/L} \cos( n \pi z/L) & n \ge 1 \end{array} \right.  \label{eq:NN1}
\end{equation}
The

\begin{figure}[htbp]
\begin{center}
\includegraphics[width=5in]{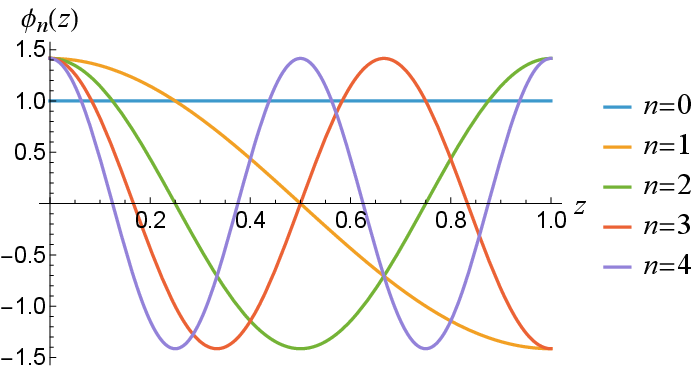}
\caption{The eigenfunctions $\phi_n(z)$, as given by Eq, (\ref{eq:NN1})}
\label{fig:Neumann-Neumann_plot}
\end{center}
\end{figure}
The interaction sum leading to the $h$ or $m$ independent contribution to the Casimir and Helmholtz force is the same as the sum  in (\ref{eq:DD6}), with the exception of the inclusion of the $n=0$ term, which is absent in the case of Dirichlet-Dirichlet boundary conditions. However, this term has no $L$-dependence and hence does not contribute to the fluctuation-induced force. 

As for the $h$-dependence of the Casimir force, a constant ordering field couples only to the uniform $n=0$ eigenfunction, as in the case of periodic boundary conditions. For this reason, the analysis proceeds in the same way for the two boundary conditions. There is no $h$ dependence in the Casimir force for Neumann-Neumann boundary conditions  and, therefore, there is no $m$ dependence in the Helmholtz force. 

\section*{Acknowledgements} DD acknowledges, that this work was accomplished by the Center of Competence for Mechatronics and Clean Technologies “Mechatronics, Innovation, Robotics, Automation and Clean Technologies” – MIRACle, with the financial support of contract No. BG16RFPR002-1.014-0019-C01, funded by the European Regional Development Fund (ERDF) through the Programme “Research, Innovation and Digitalisation for Smart Transformation” (PRIDST) 2021–2027.  DD also acknowledges the partial financial support via Grant No KP-06-H72/5 of Bulgarian NSF.

%\bibliography{publications}

	%\input publications.tex

\end{document}